\documentclass [10pt,journal,compsoc]{IEEEtran}
\usepackage{algorithm,amsbsy,amsmath,amssymb,epsfig,bbm,mathrsfs,multirow,amsthm,xurl}
\ifCLASSOPTIONcompsoc
\usepackage{cite}
\else
\usepackage{cite}
\fi
\usepackage{color}
\usepackage{algpseudocode}
\usepackage{graphicx,caption,subcaption,array,multirow,url}
\usepackage{makecell,cite}

\DeclareMathAlphabet{\mathpzc}{OT1}{pzc}{m}{it}

\newtheorem{lemma}{Lemma}
\newtheorem{theorem}{\textbf{\textsc{Theorem}}}

\usepackage{hyperref}
\hypersetup{
	colorlinks=true,
	linkcolor=blue,
	filecolor=blue,   
	urlcolor=black,citecolor=blue,}
\begin{document}
	\title{MetaShard: A Novel Sharding Blockchain Platform for Metaverse Applications}
	\author{Cong T. Nguyen, Dinh Thai Hoang, Diep N. Nguyen, Yong Xiao, Dusit Niyato, and Eryk Dutkiewicz %
		\thanks{Preliminary results of this work have been reported at the IEEE 95th Vehicular Technology Conference
			(VTC2022-Spring), Helsinki, Finland.~\cite{metachain}.
		}
\IEEEcompsocitemizethanks{
	
	\IEEEcompsocthanksitem Cong T. Nguyen, Diep N. Nguyen, Dinh Thai Hoang, and Eryk Dutkiewicz are with the School of Electrical and Data Engineering, University of Technology Sydney, Australia.	E-mail: cong.nguyen@student.uts.edu.au and  \{diep.nguyen, hoang.dinh, eryk.dutkiewicz\}@uts.edu.au.
	
	\IEEEcompsocthanksitem Yong Xiao is with the School of Electronic Information and Communications at the Huazhong University of Science and Technology, Wuhan 430074, China, also with the Peng Cheng Laboratory, Shenzhen, Guangdong 518055, China, and also with the Pazhou Laboratory (Huangpu), Guangzhou, Guangdong 510555, China (e-mail: yongxiao@hust.edu.cn).
	
	\IEEEcompsocthanksitem Dusit Niyato is with the Nanyang Technological University, Singapore 639798 (e-mail:DNIYATO@ntu.edu.sg).
}
		
		\vspace{-5mm}}
\IEEEtitleabstractindextext{	
	\begin{abstract}

Due to its security, transparency, and flexibility in verifying virtual assets, blockchain has been identified as one of the key technologies for Metaverse. Unfortunately, blockchain-based Metaverse faces serious challenges such as massive resource demands, scalability, and security/privacy concerns. To address these issues, this paper proposes a novel sharding-based blockchain framework, namely MetaShard, for Metaverse applications. Particularly, we first develop an effective consensus mechanism, namely Proof-of-Engagement, that can incentivize MUs’ data and computing resource contribution. Moreover, to improve the scalability of MetaShard, we propose an innovative sharding management scheme to maximize the network's throughput while protecting the shards from 51\% attacks. Since the optimization problem is NP-complete, we develop a hybrid approach that decomposes the problem (using the binary search method) into sub-problems that can be solved effectively by the Lagrangian method. As a result, the proposed approach can obtain solutions in polynomial time, thereby enabling flexible shard reconfiguration and reducing the risk of corruption from the adversary. Extensive numerical experiments show that, compared to the state-of-the-art commercial solvers, our proposed approach can achieve up to 66.6\% higher throughput in less than 1/30 running time. Moreover, the proposed approach can achieve global optimal solutions in most experiments. 		
		
	\end{abstract}
	
	}

\maketitle
\IEEEdisplaynontitleabstractindextext

\IEEEpeerreviewmaketitle

\thispagestyle{empty}
	
	\section{Introduction}
	Being considered as the future of Internet applications, Metaverse has recently attracted massive attention from both the industry and academia. Metaverse is commonly referred to as virtual 3D environments where humans, represented by their digital avatars, can take part in a wide range of activities such as meetings, education, gaming, and so on. Compared to traditional 3D virtual worlds, Metaverse offers users the unique ability to seamlessly move between different virtual worlds with their avatars to enjoy a wide range of services, thereby enabling much greater immersive experiences and user freedom~\cite{meta1,meta2,meta3,book}. With promising potential, Metaverse has attracted huge investments, e.g., from Meta (Facebook), Roblox~\cite{web}, Adidas~\cite{adidas}, and Microsoft~\cite{micro}. As a result, it is expected that Metaverse applications will become an increasingly important part of the future Internet and rival traditional Internet applications.
	
	However, the development of Metaverse applications has been facing several novel challenges. First, since Metaverse applications are expected to serve hundreds of millions of Metaverse users (MUs), the demands for communication and computing resources may exceed the capacity of existing digital infrastructure, e.g., 100 times more demanding in terms of computing resources~\cite{demand}. Moreover, high interoperability among different applications is necessary to allow MUs to seamlessly move between different virtual worlds. Furthermore, ensuring security and privacy for MUs in such a complex environment is a challenging task for Metaverse Service Providers (MSPs)~\cite{meta1,meta2,meta3,book}.
	
	To address those challenges, blockchain technology has been identified as one of the key technologies for Metaverse~\cite{meta1,meta2,meta3}. Particularly, blockchain technology enables a decentralized platform to securely store and manage data and complex interactions in the Metaverse. For example, with the ability to ensure data integrity, blockchain can be utilized to verify and authenticate MUs' identities, digital assets, and transactions. Moreover, due to its decentralized nature, blockchain can avoid single-point-of-failure, alleviate the heavy burden on central servers, as well as utilize resources from millions of MUs. Furthermore, blockchain can help to improve MUs' privacy, while maintaining a high level of transparency and trust for MUs. 
	
	Despite its undeniable role in the development of Metaverse, blockchain technology has several limitations. In particular, traditional blockchain solutions based on Proof-of-Work (PoW), are usually very slow in processing with high-demand of computing resource~\cite{wangsurvey,Xiaosurvey}, which makes them unsuitable for Metaverse applications. Unfortunately, Metaverse applications may require high levels of scalability to support a huge number of MUs and transactions, which is beyond the capacity of conventional blockchain technology~\cite{wangsurvey,Xiaosurvey}. Furthermore, interoperability is another critical issue that blockchain technology is facing. Specifically, different blockchain networks and protocols are often incompatible with each other, making it difficult for them to exchange data and information~\cite{fedchain}. 
	
	Several recent efforts have been made to address those challenges. Particularly, recent blockchain networks have been employing the Proof-of-Stake (PoS) consensus mechanism which replaces the compute-intensive puzzle-solving process of PoW with the stake ownership requirement. As a result, PoS has higher transaction processing capabilities while consuming negligible computing resources, compared to the PoW-based consensus mechanisms~\cite{wangsurvey,Xiaosurvey,PoS}. However, despite its outstanding benefits, conventional PoS consensus mechanisms' transaction processing capabilities are still inadequate to meet the huge demands of Metaverse applications. To address this issue and improve the scalability of the solution, sharding mechanism~\cite{shardsurvey1,shardsurvey2} has been recently developed to divide the blockchain network into multiple sub-networks (shards). Each shard can process transactions independently and in parallel to other shards, thereby significantly improving the transaction processing speed. There are, however, trade-offs between security and speed, i.e., the more shards a network is divided into, the less secure the network becomes. Particularly, dividing the blockchain into multiple shards makes it easier for the adversary to conduct 51\% attacks~\cite{shardsec}. For example, if a PoS-based blockchain network is divided into 10 shards, the adversary might only need 6\% of the network total stakes (coins) to successfully perform a 51\% attack, whereas it will need at least 51\% of stakes if there is no shard in the network. Therefore, it is crucial to determine the number of shards as well as the MU allocation in each shard to ensure that the adversary cannot attack any shard in the network. Unfortunately, this problem has not been well investigated in the literature (as discussed in more detail in Section~\ref{sec:rel}). Furthermore, these approaches, e.g., PoS and sharding, cannot address the massive resource demands of Metaverse applications. Therefore, an intelligent blockchain framework that can address these challenges and at the same time meet the high resource demands of Metaverse is in urgent need.  
	
	Motivated by the above, we develop MetaShard, a novel sharding-based blockchain framework that can not only leverage MUs resources contributions to alleviate the burdens on the MSP but also improve scalability while ensuring the security of the whole system. To this end, we first develop an innovative consensus mechanism, namely Proof-of-Engagement (PoE), that can incentivize MUs to participate in the consensus process and contribute computing resources and/or collected IoT data to the Metaverse applications. Based on their engagement (determined by their contributions and assets), MUs can be rewarded with blockchain tokens via the block reward. As a result, PoE can utilize MUs resources to alleviate the huge resources burden for MSP and create a more engaged MUs community. Moreover, to improve scalability, we propose a sharding management scheme to divide the network into multiple shards to enable parallel transaction processing, thereby significantly improving the network throughput. Furthermore, to protect the shards from 51\% attack, we formulate an optimization problem to find the optimal number of shards and MUs allocation, thereby maximizing the network throughput while ensuring that the risk of attacks in individual shards is minimal. Since the problem is NP-complete, we develop a hybrid approach that first decomposes the problem using binary search and then solves the relaxed sub-problems using Lagrange multipliers. As a result, the proposed approach can quickly obtain solutions, improve the network performance, and at the
	same time enhance security compared to those of state-of-the-
	art solvers. The main contributions of this paper can be summarized as follows:
	\begin{itemize}
		\item We propose MetaShard, a novel sharding blockchain framework for Metaverse applications that not only leverage MUs' resources and data for Metaverse applications but also enhance network throughput. 
		\item We develop PoE, a new consensus mechanism that can incentivize MUs' data and computing resources contribution via the block rewards, thereby alleviating the massive resource demands and incentivizing MUs to be more engaged in the Metaverse.
		\item  We propose a sharding management scheme to improve the scalability of MetaShard. Specifically, we first formulate an optimization problem to maximize the network throughput while protecting the shards from 51\% attacks. We then develop a lightweight hybrid approach to quickly obtain solutions, thereby allowing flexible shard reconfiguration.
		\item We conduct extensive simulations to evaluate the performance of our proposed approach. The results show that, compared to the state-of-the-art commercial solvers, our proposed approach can achieve up to 66.6\% higher throughput in less than 1/30 running time. Moreover, the proposed approach can achieve global optimal solutions in most experiments. Furthermore, we study the impacts of key parameters on the performance of the system and show that the proposed approach can further improve the robustness of the system. 
	\end{itemize}     

	The rest of the paper is organized as follows. The related
	work is discussed in Section~\ref{sec:rel}. Section~\ref{sec:SM} presents MetaShard's system	overview. The proposed PoE consensus mechanism and sharding management scheme are presented in detail in Section~\ref{sec:con}. Section~\ref{sec:shard} presents the sharding management problem and our proposed lightweight approach. Finally, Section~\ref{simu} shows the system performance and Section~\ref{conclu} concludes the paper.

\section{Related Work}
	\label{sec:rel}
	\subsection{Blockchain for Metaverse}
	
	As Metaverse is an emerging topic, applications of blockchain in Metaverse are still very limited. There are just a few recent works~\cite{s1,s2,s3} focusing on this topic. Specifically, in~\cite{s1}, the authors propose a blockchain-based secure mutual authentication scheme for Metaverse environments. In this approach, the MUs need to send their pseudo-identity, personal information, and public key to a central authority to verify. If the verification is successful, the central authority stores the MUs' identities and public keys in a public blockchain for Metaverse applications to query. Similarly, the authors in~\cite{s3} develop a blockchain-based framework for Metaverse to manage MUs' identities and transactions. Particularly, the proposed framework is composed of four parts, namely New User Engine, Transaction Centre,	Authenticator Engine, and Repo. In this framework, the New User Engine is responsible to provide new MUs with blockchain addresses. MUs can then send their transactions to the Transaction Centre to process, and the Authenticator Engine's responsibility is to validate the MUs' identities and transactions. If the transaction is successfully validated, it will be recorded in the Repo (which is a distributed ledger) along with the resulting change in MUs' accounts.  In~\cite{s2}, the authors propose a blockchain-enabled framework for Metaverse service management. Particularly, in the proposed framework, the mobile network operators can offer their services to MUs with different service level agreements and prices. The MUs can then choose one of the options based on a proposed utility function with a trade-off between service quality and prices. In this framework, the blockchain serves as a platform to verify MUs' identities, and the blockchain tokens are used as the currency for payment.
	
	From the above, we can observe that~\cite{s1,s2,s3} only utilize conventional blockchain technology for managing MUs identities and transactions without taking into account specific challenges of Metaverse, such as the huge resource demand or the associated scalability issues. To the best of our knowledge, our proposed MetaShard framework is the first in the literature that can encourage MUs to contribute resources to the Metaverse and blockchain network as well as address the scalability issue of blockchain. 
	\subsection{Sharding in Blockchain}
	In~\cite{rapidchain}, the authors propose a sharding protocol for public blockchain networks. Although the protocol is proven to be secure, it utilizes PoW to authenticate the consensus participants' identities. Another PoW-based sharding scheme is proposed in~\cite{jsac}, where nodes with high computing power in the system can participate in several shards simultaneously. Similar to~\cite{rapidchain}, this scheme requires consensus participants to solve a PoW puzzle to become validators, and shards' security is proven based solely on the number of consensus participants. However, since the consensus participants are required to solve a PoW puzzle, the adversary can split their computing power to simultaneously solve different puzzles and thus able to gain more slots. As a result, the computing power distribution needs to be taken into account, but it is not discussed in~\cite{rapidchain} and~\cite{jsac}. Another PoW-based sharding scheme is proposed in~\cite{elastico}. Although the scheme's security is proven, it relies on PoW, which is unsuitable for Metaverse due to the high delay and huge energy consumption. 
	
	To address the limitations of PoW, other sharding protocols were developed with energy-saving alternative ways to select consensus participants. For example, in~\cite{shard2}, a sharding protocol is developed based on Byzantine Fault Tolerance (BFT) and Trusted Execution Environment (TEE). Particularly, the consensus participants need a special type of hardware to ensure the TEE. A similar approach that relies on TEE is proposed in~\cite{new1}. Particularly, a sharding scheme is developed that utilizes two separate blockchains to decouple the transaction recording and consensus processes. Similar to~\cite{shard2}, the proposed scheme relies on TEE, and thus it also requires special hardware to participate in the consensus process. Although the schemes in~\cite{shard2} and~\cite{new1} can enhance the security of the network, the hardware requirement makes them much less attractive to public users, especially MUs who already need a lot of computing power for AR/VR rendering. In~\cite{shard3}, the authors develop a sharding scheme based on Practical Byzantine Fault Tolerance (PBFT). Although the security of the protocol is proven, how to select the consensus participants is not discussed. Moreover, similar to~\cite{rapidchain}, this protocol relies on the number of consensus participants, without taking into account the ability of the adversary to create many identities to gain more consensus participants slots. In~\cite{shard4}, a reputation-based sharding scheme is developed. Particularly, the consensus participants are selected based on their reputation scores stored in a separate blockchain. Then, the selected consensus participants execute a BFT-based protocol for each shard's consensus process. However, the adversary in this case can also create many identities to increase the number of consensus participants it controls, as the reputation is based solely on previous behaviors. In~\cite{Omni}, a BFT-based sharding protocol is developed. However, similar to~\cite{rapidchain} and~\cite{shard4}, the protocol relies on the number of consensus participants, which can be adversely impacted by the adversary. In~\cite{shard5}, a dynamic sharding protocol is proposed in which the consensus participants are selected via smart contracts. Moreover, to mitigate Sybil attacks, the proposed protocol requires that each consensus participant must come from a different IP address. Nevertheless, this still cannot prevent the adversary from influencing the selection process, as IP addresses can be fake. 
	
	From the abovementioned approaches, we can observe that they rely on the PoW consensus mechanism which is inappropriate for Metaverse due to the huge energy consumption and large delay. In contrast, our proposed PoE consensus mechanism is much more energy-efficient. Moreover, the security of these approaches relies on the number of consensus participants without considering that this number can be unfairly affected by the adversary. On the contrary, our proposed approach considers the MUs' engagement instead of the number of participants, thereby enhancing the security and robustness of the system against Sybil attacks.
	
	\section{System Overview}
	\label{sec:SM}

\subsection{System Overview}
		\begin{figure}[!]
	\includegraphics[width=.5\textwidth]{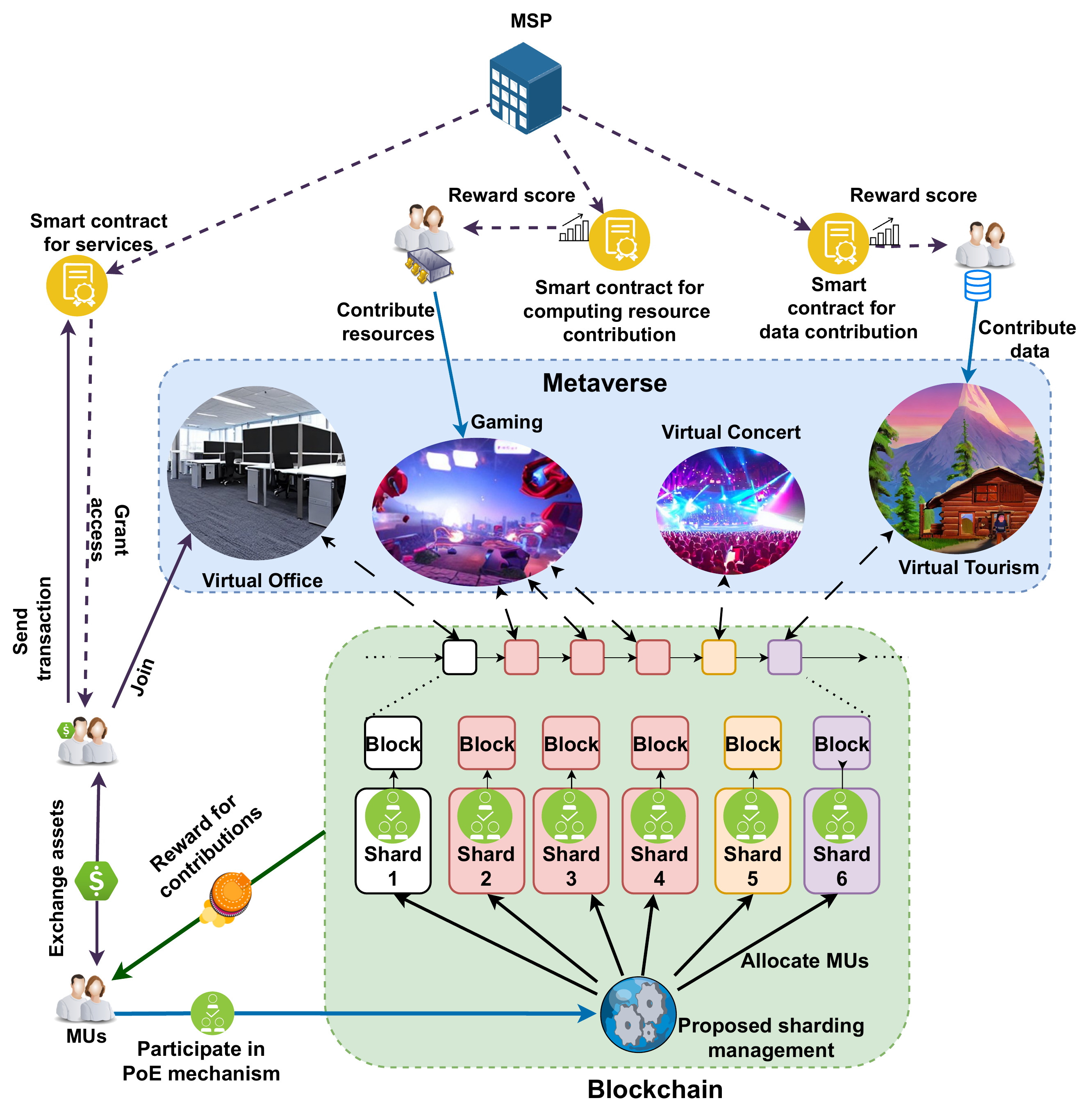}
	\centering
	\caption{An illustration of the proposed system.}
	\label{Fig:system}
\end{figure}
Fig.~\ref{Fig:system} illustrates an overview of the proposed MetaShard framework. In this framework, there is an MSP operating a Metaverse running multiple Metaverse applications. Each Metaverse application is a self-contained environment that offers a wide range of services and experiences, e.g., virtual office, virtual concerts, gaming, and virtual tourism, for MUs. Compared to traditional virtual applications, the core difference here is that applications in the Metaverse are fully interconnected, allowing the MUs to freely and seamlessly move between different applications, e.g., Meta Horizon Worlds~\cite{horizon}. The MUs also have various interactions with each other and the MSP, such as exchanging assets, purchasing services and items, contributing resources, and participating in the blockchain's consensus process. A blockchain-based system can be applied to record and facilitate those interactions. 
\subsection{Metaverse Users (MUs) and Metaverse Service Provider (MSP)}
An MU is a user that can join and use different Metaverse applications and services provided by the MSP. The MUs have unique avatars that represent them in the Metaverse, allowing them to interact with each other as well as the virtual worlds. There can be various interactions among the MUs, as well as between the MUs and the MSP. First, the MUs can easily exchange digital assets, such as Metaverse tokens and virtual items, with each other using blockchain transactions. For example, MUs who purchased virtual concert tickets (but could not attend) can sell their tickets to others. All these digital assets and transactions can be verified and stored in the blockchain, providing a secure transparent way to manage assets without the need for a central authority.

Moreover, the MUs can pay the MSP to gain access to services or buy digital items. This process can be automated by smart contracts, i.e., a user-defined program that can be automatically executed when the
conditions within are met~\cite{SC}. For example, the MSP can broadcast its virtual meeting options, e.g., duration, number of people, and fees, by publishing a smart contract on the blockchain. Then, MUs who want to purchase this service can send a transaction that contains the specified options to the smart contract. After the transaction is validated, the smart contract can automatically send the MU a transaction that contains a proof for the purchase. When the MU requests to enter the virtual meeting room, the Metaverse application can query the blockchain to verify the proof and grant the involved MUs access to the room.  

Furthermore, in our proposed MetaShard, MUs can also contribute data or computing resources to Metaverse applications. For example, in Metaverse virtual tourism applications, the MSP needs up-to-date 3D image/video data from tourist attractions to provide more immersive experiences to MUs. In this scenario, the MSP can encourage MUs who live near the tourist attraction to contribute the data, thereby saving costs and increasing MUs' engagement. Moreover, in compute-intensive AR/VR applications, the MSP can incentivize MUs to execute the rendering locally instead of offloading to the MSP's servers. Additionally, the MSP can offload computing tasks to MUs with idle resources to alleviate the heavy burden on the edge/cloud servers. For those contributions, MUs can be rewarded with digital assets such as Metaverse items or tokens. This can help to encourage more MUs to participate in the Metaverse and alleviate the high resource demands of Metaverse applications. Similarly, smart contracts can be utilized to provide a transparent and trusted way to reward the MUs for their contributions because the conditions written within a smart contract are visible to everyone. For example, the MSP can publish a smart contract that specifies the payment for different amounts of data contributed. When the MUs send the data to the smart contract, they can be automatically paid for their data. 
\subsection{Blockchain and Sharding}
In MetaShard, the blockchain serves as a platform to store and manage MUs and applications data, interactions, and assets. Blockchain enables the MUs and the MSP to manage their identities, avatars, and digital assets in a decentralized manner, thereby significantly enhancing transparency and trust for MUs. Moreover, smart contracts can automate and facilitate various interactions among MUs and applications. Furthermore, the blockchain can also provide a transparent way to manage and reward MUs' data and computing resources contribution, thereby creating a more engaged and motivated MUs community. However, managing such a huge amount of data and interactions for many MUs requires very high transaction processing capabilities, which conventional blockchain technology cannot handle. Particularly, most current blockchain networks are still employing the PoW consensus mechanism which consumes a huge amount of energy and has very low processing capability. 

Therefore, we propose a PoS-based consensus mechanism for MetaShard. With PoS, the energy consumption is negligible, and the transaction processing capability can be significantly improved. Moreover, different from the conventional PoS that only considers the user assets (stakes), we develop a PoE consensus mechanism that will also take into account MUs' data and resources contribution and reward MUs for their engagement. In this way, PoE can not only leverage MUs' resources to alleviate the massive resource demands for the MSP but also encourage more MUs to join the Metaverse for the rewards, thereby creating a more engaged MUs community. This PoE consensus mechanism will be discussed in detail in Section~\ref{sec:con}. Moreover, scalability is a major constraint that hinders the applicability of conventional blockchain technology for Metaverse applications with a huge number of MUs. Therefore, we propose to employ the sharding mechanism~\cite{shardsurvey1,shardsurvey2} for MetaShard. With sharding, the blockchain network can be divided into multiple smaller networks that allow the parallel processing of transactions and smart contracts, thereby improving scalability and processing speed and reducing the workload on individual consensus nodes. Furthermore, each Metaverse application can be adaptively allocated a different number of shards according to their processing demands. For example, we can allocate more shards to virtual office applications during working hours and more shards to virtual concert applications at night. 

Although dividing a blockchain into shards can significantly enhance the network throughput (in terms of the number of transactions successfully verified and processed per time unit), it also causes some potential risks for network security as shown in~\cite{shardsurvey1,shardsurvey2}. Particularly, the security of a blockchain network depends on the honest majority. For example, if the adversary can control the majority (51\%) of stakes in PoS, it can successfully perform various attacks, such as double-spending and transaction denial attacks~\cite{wangsurvey,Xiaosurvey,PoS}, on the network. However, if the stakes are not allocated properly into the shards, then the adversary may not need too many stakes to successfully attack a shard. Therefore, it is crucial to determine the proper number of shards and MUs allocation such that the security of the whole network is still ensured. To this end, in Section~\ref{sec:shard}, we will formulate this sharding management optimization problem and propose an efficient approach to quickly obtain solutions, thereby significantly improving the network performance and security. 
\section{Proposed PoE Consensus Mechanism and Sharding}
\label{sec:con}
\subsection{Epoch and Time Slots}
		\begin{figure}[!t]
	\includegraphics[width=.5\textwidth]{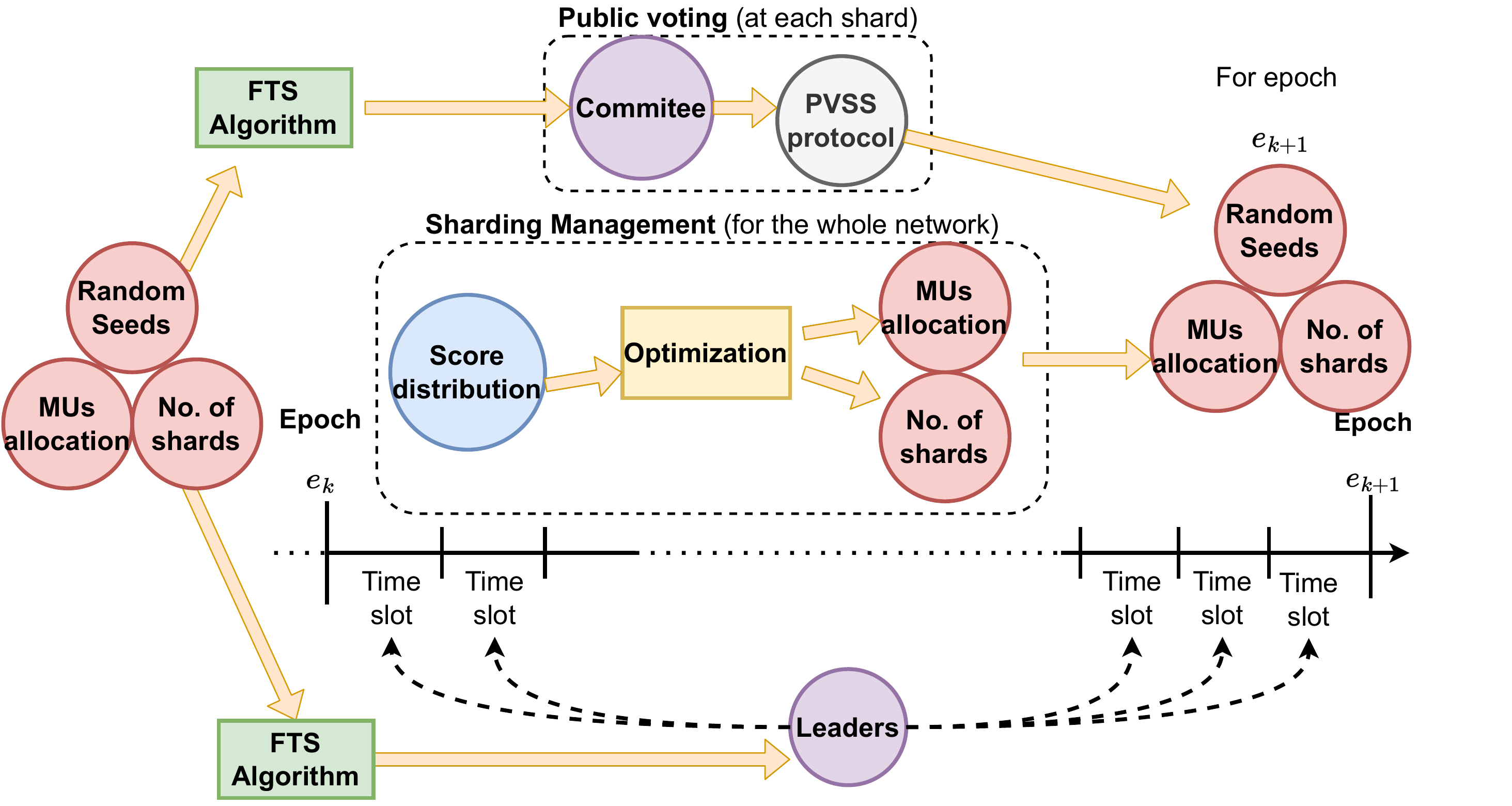}
	\centering
	\caption{An illustration of the proposed sharding management and election processes.}
	\label{Fig:epoch}
\end{figure}
In our proposed PoE consensus mechanism, time is divided into
epochs. Each epoch is then divided into time slots. During epoch $e_k$, our proposed sharding management process is executed to determine the number of shards and MUs allocation for epoch $e_{k+1}$, as illustrated in Fig.~\ref{Fig:epoch}. Note that frequent and dynamic adjustment of the number of shards can be beneficial for the system, e.g., adding more shards to address the varying transaction processing demands or closing shards to reduce unnecessary communication~\cite{dynamic}. This sharding management process is run once for each epoch, which is beneficial for network security~\cite{shardsurvey1,shardsurvey2}. Moreover, it is also more desirable for the MUs, e.g., an MU who contributes more in this epoch should have a higher chance to be elected as a leader and earn block rewards (e.g., in Metaverse tokens) in the next epoch. 

Moreover, during the epoch, the committee members (selected from MUs who participate in the consensus process) of each shard execute the Publicly Verifiable Secret Sharing
(PVSS) protocol~\cite{PVSS} to create random seeds. The PVSS protocol is guaranteed to produce unbiased random strings, and it allows network participants to verify those strings, as long as 51\% of the protocol participants are honest~\cite{PVSS}. Therefore, the PVSS protocol can be employed to create publicly verifiable random seeds. At the beginning of each epoch, these random seeds, along with the number of shards and MUs allocation, are then used as the input of a hash function, e.g., Follow-the-Satoshi (FTS) algorithm~\cite{PoS}, to choose the leaders for the current epoch and committee members for the next epoch. If the numbers of shards of two epochs are different, the random seeds can be used to determine which shard will create more (or fewer) random seeds. For example, if there is one more shard in the next epoch, then a random shard in this epoch will create two seeds instead of one.



\subsection{MU Engagement and Reward}
In MetaShard, MUs are incentivized to contribute data and computing resources. To reward this contribution, MUs are given contribution scores that are stored in the blockchain. These scores are then used along with the MUs assets, e.g., Metaverse items and tokens, to determine the MUs' total engagement scores. Particularly, each MU has a data contribution score $D_n$, a computing resource contribution score $C_n$, and an amount of Metaverse tokens $T_n$. The data and computing resource score rewarded to the MUs can be determined by the MSP, e.g., based on the amount or frequency of resources and data contribution~\cite{contribution,contribution2}. The total engagement score of MU $n$ can be calculated by
\begin{equation}
	\eta_n=\alpha_D D_n+ \alpha_C C_n+\alpha_T T_n,
\end{equation}
where $\alpha_D,\alpha_C$, and $\alpha_D$ are the weight factors for data contribution, computing resources contribution, and Metaverse token, respectively. These weight factors are determined by the MSP, and they can also reflect the MSP's priority. For example, if the MSP needs more computing resource contribution, it can set $\alpha_C$ higher than $\alpha_T$ and $\alpha_D$.

Every MU can choose to participate in the consensus processes to be able to earn the block rewards. Since each shard runs its own consensus process, the probability that MU $n$ is selected to be the leader of shard $s$ is given by:
\begin{equation}
	\label{eq1}
	\rm{Pr}^s_n=\dfrac{\eta_n^s}{\sum_{i=1}^N \eta_i^s}.
\end{equation}
Besides the benefits of MUs' resource contribution, our proposed leader selection approach can also enhance the security of the network. The reason is that MUs who are more engaged (with high contributions and own a lot of assets) might want to protect the network more. Moreover, in existing approaches such as~\cite{rapidchain,Omni,jsac,new1,elastico,shard3,shard4,shard5}, the leader is not selected based on stakes/scores (BFT-based approaches). Instead, these approaches rely only on the number of validators. However, the adversary can target those protocols by conducting Sybil attacks, i.e., creating multiple accounts, to improve their chance of being selected as validators. In contrast, the leaders are chosen based on their engagement in MetaShard, and thus creating multiple accounts with no contributions or assets cannot adversely affect the leader selection process.

\subsection{Threat Model and Shard Security}
\textbf{Threat Model:} In this work, we consider the type of adversary that tries to gain the majority in any shard to conduct 51\% attacks. Particularly, the adversary possesses multiple accounts (adversarial MUs) in the system. These accounts, along with the other MUs' accounts, are allocated into different shards in the system. If the total score of the adversary exceeds 51\% of the total score of any shard in the system, the adversary can successfully conduct various attacks, such as double-spending and transaction denial attacks~\cite{wangsurvey,Xiaosurvey,PoS}, and unfairly affect the seeds generation of the PVSS protocol.
Moreover, the adversary can corrupt honest MUs, but the corruption will take effect after a period of time~\cite{elastico,rapidchain,Omni,shardsurvey1}. When an MU is corrupted, it will be controlled by the adversary, and its score will count toward the adversary's total score. 

Given the above adversary model, there are two serious threats. First, when the adversary controls more than 51\% of a shard, the adversary can influence the leader election process to conduct other types of attacks such as double-spending and transaction denial attacks~\cite{wangsurvey,Xiaosurvey,PoS} on the shard. Consequently, the Metaverse transactions might be reverted, or transactions from specific MUs might be blocked by the adversary. Therefore, it is crucial to allocate scores to each shard such that the adversary has a minimal chance to attack every shard. However, a major challenge is that we do not know which MU is adversarial, and thus we can only minimize the chance that the adversary can control the majority of scores in any shard. Second, if the epoch is too long, the adversary might be able to corrupt the honest MUs during the epoch and successfully gain control of the shard. Therefore, the score allocation needs to be regularly reconfigured, e.g., Ethereum's epoch only lasts for 6.4 minutes~\cite{Eth}. 

To address these threats, we develop a sharding management approach to determine the number of shards and allocate MUs scores such that the adversary's chance to successfully attack any shard is minimal, e.g., lower than 0.1\%. Moreover, the proposed approach can quickly obtain solutions, thereby reducing the time for the adversary to corrupt honest MUs. The proposed approach is presented in detail in the next section.

\section{Sharding Management Problem and Solution}
\label{sec:shard}
\subsection{Problem Formulation}

We first formulate the sharding management problem as follows. In the considered system, there is a set $\mathcal{N}=(1,\ldots,N)$ of MUs. Since we do not know which MU is adversarial, we can consider the total engagement score of the adversary, denoted by $\eta^A_s$, to be a sum of independent random variables. Let $p^A_n$ denote the probability that MU $n$ is adversarial. $p^A_n$ can be determined based on the MUs' assets and contribution, i.e., MUs who owns more assets or contribute frequently to the Metavese are less likely to be adversarial, or using Machine Learning approaches such as those in~\cite{predict1,predict2,predict3}. The expected value of the total engagement score of the adversary in shard $s$ can then be determined by:
\begin{equation}
	\label{adv_stake}
\mathbb{E}[\eta^A_s]=\mathbb{E}[\sum_{n=1}^N p^A\eta^s_n]=\sum_{n=1}^N p^A \eta^s_n.
\end{equation} 
Since $\eta^A_s$ is a sum of independent random variables, we want to determine the probability that $\eta^A_s$ exceeds 50\% of the total engagement scores in any shard, i.e., when the adversary gains the majority in a shard. To find this probability, we apply the Hoeffding bound~\cite{stat} to determine the bounds on the tail distribution of $\eta^A_s$ . Particularly, let $\theta_s=\sum_{n=1}^N \eta^s_n$ denote the total engagement score of all MUs (including the adversary) in shard $s$. Based on~\eqref{adv_stake}, the probability that the adversary's score exceeds 50\% of the total scores in shard $s$ can be determined by:
\begin{equation}
	\label{prob_1}
	\begin{split}
		\textrm{Pr} [\eta^A_s \geq 0.5\theta_s]&=\textrm{Pr} [\eta^A_s \geq \mathbb{E}[\eta^A_s]+t]\leq \textrm{exp}\big(\dfrac{-2 t^2}{\sum_{n=1}^{N} (\eta_n^s)^2}\big)
	\end{split}
\end{equation}
where $t$ denotes the deviation from the expected value of $\eta^A_s$ such that the adversary can gain majority in the shard, i.e., $\eta^A_s\geq 0.5\theta_s$. This deviation can be determined by:
\begin{equation}
	\begin{split}
		0.5\theta_s=\mathbb{E}[\eta^A_s]+t,\\
		\sum_{n=1}^N  0.5\eta^s_n=t+\sum_{n=1}^N  p^A_n\eta^s_n, \\
		t=\sum_{n=1}^N (0.5-p^A_n) \eta^s_n.
	\end{split}
\end{equation}
The inequality in~\eqref{prob_1} comes from Hoeffding bound~\cite{stat}.
To keep the probability in~\eqref{prob_1} lower than a certain safety threshold $\tau$ (e.g., $\tau=0.001$), we have
\begin{equation}
	\label{threshold}
	\begin{split}
		\text{exp}(\dfrac{-2t^2}{\sum_{n=1}^{N} (\eta_n^s)^2})\leq \tau,\\
		-2\dfrac{\bigg(\sum_{n=1}^N (0.5-p^A_n) \eta^s_n\bigg)^2}{\sum_{n=1}^{N} (\eta_n^s)^2}\leq \ln(\tau),\\
		(\sum_{n=1}^N (0.5-p^A_n) \eta^s_n)^2\geq -0.5\ln(\tau)\sum_{n=1}^{N} (\eta_n^s)^2.\\
	\end{split}
\end{equation}
This means that to make all the shards to be secured, we need to allocate the scores $\eta_n^s$ of the MUs in each shard such that they satisfy~\eqref{threshold}. Let $S$ denote the maximum possible number of shards\footnote{In theory, we do not have the maximum possible number of shards, e.g., an MU can participate in many shards. However, in practice, this number cannot be unlimited because an MU does not want to participate in too many shards (same rewards but needs much more computational and communication resources).}. We formulate the optimal sharding management problem \textbf{(P1)} below.

\begin{align}
	(\textbf{P1})&\max_{\boldsymbol{\eta},\mathbf{x},\varsigma} T\varsigma&\label{obj}\\
	\textrm{s.t.}& \hspace{0.5 em}  (\sum_{n=1}^N (0.5-p^A_n) \eta^s_n)^2\geq -0.5x_s\ln(\tau)\sum_{n=1}^{N} (\eta_n^s)^2,&\nonumber\\ 
	&\hspace{14 em}\forall s=1,\ldots,S \label{c1} \\
	&x_s \geq \dfrac{\varsigma-s+1}{S},\hspace{7.5 em}\forall s=1,\ldots, S\label{c2}\\
	&x_s \leq \varsigma-s+1, \hspace{7.5 em}\forall s=1,\ldots, S\label{c3}&\\
	&\sum_{s=1}^{S} \eta_n^s =\eta_n,\hspace{8.5 em}\forall n \in \mathcal{N}.	 \label{c4}
\end{align}
In (\textbf{P1}), the objective~\eqref{obj} is to maximize the total network throughput, which can be obtained by multiplying the number of shards $\varsigma$ with the maximum number of transactions that a shard can process per second $T$. Constraints~\eqref{c1} follow~\eqref{threshold}. Note that out of these $S$ constraints, only $\varsigma$ constraints are active to ensure the security for $\varsigma$ shards, while the constraints for the other (dummy) shards need to be inactive. To this end, we use auxiliary decision variables $\mathbf{x}$ to make the constraints active for the shards from 1 to $\varsigma$, and inactive for the other shards. Particularly, constraints~\eqref{c2} and~\eqref{c3} ensure that $x_s=1, \forall s=1,\ldots, \varsigma$, while $x_s=0, \forall s=\varsigma,\ldots, S$. Then, for shards from 1 to $\varsigma$, the right-hand-side of constraints~\eqref{c1} become $-0.5x_s\ln(\tau)\sum_{n=1}^{N} (\eta_n^s)^2$ (active). For shards from $\varsigma+1$ to $S$, the right-hand-side of constraints~\eqref{c1} become zero, and thus they are always satisfied (inactive). Finally, constraints~\eqref{c4} ensure that the MUs scores are fully allocated. The reason for these constraints is that the MUs' rewards for consensus participation are proportional to their engagement scores, and thus the MUs will want to use all their scores for consensus participation.

From~\eqref{c1}, we can observe that (\textbf{P1}) is a Mixed Integer Non-linear Programming (MINLP) problem~\cite{MILP,solver} which is NP-complete~\cite{karp}. As later shown in Section~\ref{simu}, commercial solvers such as CPLEX~\cite{solver} can only solve instances of (\textbf{P1}) with a small number of shards. For larger values of $S$, it becomes intractable and infeasible to obtain optimal solutions. However, the score allocation needs to be regularly reconfigured, e.g., Ethereum's epoch only lasts for 6.4 minutes~\cite{Eth}. Such frequent shard reconfiguration can bring various benefits. First, the MUs who contribute more resources in one epoch can have their scores updated earlier to earn more rewards in the next epoch. Moreover, if the epoch is short, the adversary will have less time to corrupt the honest MUs. 
\subsection{Proposed Hybrid Algorithm}
\subsubsection{Problem decomposition and the proposed Lagrangian approach }
To address the abovementioned problems, we develop a lightweight approach based on Lagrange multipliers and binary search that can quickly obtain solutions in a very short time, thereby enabling flexible scores reallocation and improving the shards' security. To that end, we first decompose (\textbf{P1}) into multiple relaxed sub-problems (\textbf{P2}) as follows:

\begin{align}
	(\textbf{P2})&\max_{\boldsymbol{\eta}} T\sigma&\label{p2obj}\\
	\textrm{s.t.}& \hspace{0.5 em}  \big(\sum_{n=1}^N (0.5-p^A_n) \eta^s_n\big)^2\geq -0.5\ln(\tau)\sum_{n=1}^{N} (\eta_n^s)^2,&\nonumber\\ 
	&\hspace{14 em}\forall s=1,\ldots,\sigma \label{p2c1} \\
	&\sum_{s=1}^{\sigma} \eta_n^s =\eta_n,\hspace{8.5 em}\forall n \in \mathcal{N}	 \label{p2c2}
\end{align}
Particularly, in (\textbf{P2}), we fix the value of $\varsigma=\sigma$. In this way, we do not need to determine $\varsigma$ and $\mathbf{x}$, and thus constraints~\eqref{c1} become constraints~\eqref{p2c1}. Moreover, constraints~\eqref{c2} and~\eqref{c3} can be omitted. Furthermore, the objective function~\eqref{p2obj} becomes a constant, and thus we only need to find a feasible solution to (\textbf{P2}), instead of optimizing it. As a result, (\textbf{P2}) becomes a Nonlinear Programming (NLP) problem, which is easier to solve compared to MINLP problems~\cite{MILP,solver}. Then, we can solve (\textbf{P2}) for all values of $\sigma=1,\ldots, S$, and the largest value of $\sigma$ for which we can find a feasible solution is the global optimal solution of (\textbf{P1}). Nevertheless, (\textbf{P2}) is non-convex and nonlinear due to~\eqref{p2c1}, and thus it still requires exponential time to solve~\cite{complex1}, as later shown in Section~\ref{simu}. 

To address this limitation, we reformulate the optimization problem (\textbf{P3}) as follows:  
\begin{align}
	(\textbf{P3})&\max_{\boldsymbol{\eta}} \sum_{s=1}^{\sigma}\bigg(\big(\sum_{n=1}^N (0.5-p^A_n) \eta^s_n\big)^2+0.5\ln(\tau)\sum_{n=1}^{N} (\eta_n^s)^2\bigg)&\label{p3obj}\\
	\textrm{s.t.}& \hspace{0.5 em}  
	\sum_{s=1}^{\sigma} \eta_n^s =\eta_n,\hspace{8.5 em}\forall n \in \mathcal{N}	 \label{p3c1}
\end{align}
The core idea of (\textbf{P3}) is that, instead of finding feasible solutions that satisfy~\eqref{p2c1} and~\eqref{p2c2}, we try to maximize the left-hand-side of~\eqref{p2c1}, subject to~\eqref{p2c2}. Then, we can check the optimal solution $\boldsymbol{\eta}'$ obtained from (\textbf{P3}). 
Then, we adopt the Lagrange multipliers method to solve (\textbf{P3}) as follow. We first define the Lagrange function:
\begin{equation}
	\begin{split}
\mathcal{L}(\boldsymbol{\eta},\boldsymbol{\lambda})&=f(\boldsymbol{\eta})-\boldsymbol{\lambda}g(\boldsymbol{\eta}),\\
&= \sum_{s=1}^{\sigma}\bigg(\big(\sum_{n=1}^N (0.5-p^A_n) \eta^s_n\big)^2+0.5\ln(\tau)\sum_{n=1}^{N} (\eta_n^s)^2\bigg)\\
&-\sum_{j=1}^{N} \lambda_j\bigg(\sum_{s=1}^{\sigma}\eta_n^s-\eta_n\bigg ).
	\end{split}
\end{equation}
Then, we solve the following set of equations:
\begin{equation}
\nabla_{\boldsymbol{\eta},\boldsymbol{\lambda}}\mathcal{L}(\boldsymbol{\eta},\boldsymbol{\lambda})=0,
\end{equation}
which is equivalent to
\begin{equation}
	\label{systemeq}
	\begin{split}
		\lambda_k+\ln({\tau})\sum_{n=1}^{N} (\eta_n^s)+2p^A_k\sum_{n=1}^N (0.5-p^A_n) \eta^s_n=0,\\
		\forall k \in \mathcal{N}, \forall s= 1,\ldots, \sigma,\\
		\sum_{s=1}^{\sigma} \eta_n^s -\eta_n=0,\forall n \in \mathcal{N}.
	\end{split}
\end{equation}
Instead of solving the NLP problem (\textbf{P2}), we only need to solve~\eqref{systemeq} which is a set of $(\sigma+1)N$ equations with $(\sigma+1)N$ variables. Moreover, in~\eqref{systemeq}, all the equations are linear, and thus it is a system of linear equations. As a result, this system of equations can be solved effectively in a very short period of time compared to (\textbf{P2}). 

\begin{figure}[!]
	\includegraphics[width=.5\textwidth]{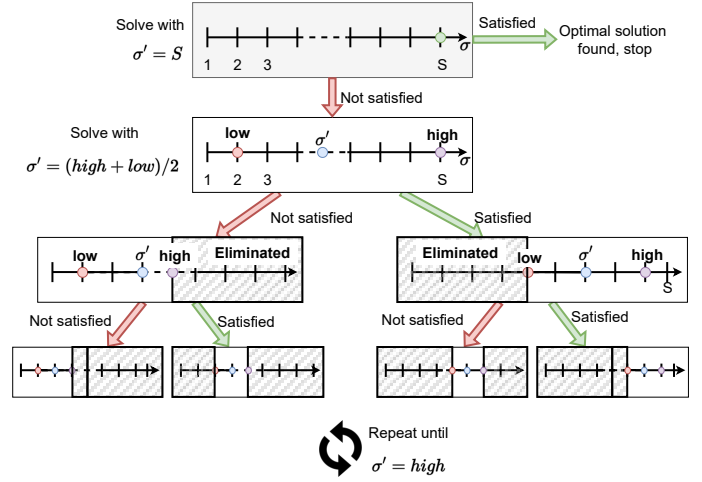}
	\centering
	\caption{An illustration of Algorithm 1.}
	\label{Fig:binary}
\end{figure}

Finally, we implement Algorithm 1 which combines binary search and the Lagrange multiplier method to obtain optimal solutions for the original problem (\textbf{P1}). Particularly, Algorithm 1 first finds the optimal solution $\boldsymbol{\eta}'$ of (\textbf{P3}), using the system of equations in~\eqref{systemeq}, with $\sigma=S$. Then, if $\boldsymbol{\eta}'$ satisfies~\eqref{p2c1}, it is the optimal solution of (\textbf{P1}). Otherwise, we apply binary search to speed up the optimization process as illustrated in Fig.~\ref{Fig:binary}. Particularly, we first set $high=S-1$, $low=2$, and $\sigma'=(high+low)/2$. Then, we solve~\eqref{systemeq} to find $\boldsymbol{\eta}'$. Next, if $\boldsymbol{\eta}'$ satisfies~\eqref{p2c1} (which means $\sigma$ is the best solution so far), we set $high=S-1$, $low=\sigma$, and $\sigma'=(high+low)/2$. Otherwise, we set $high=\sigma$, $low=2$, and $\sigma'=(high+low)/2$. In both cases, the loop is repeated until $\sigma'=high$. During the loop, the algorithm records the best solution found (that can satisfy~\eqref{p2c1}) in $\boldsymbol{\eta}^*$ and $\sigma^*$, and it will return $\boldsymbol{\eta}^*$ when the loop ends. With $\boldsymbol{\eta}^*$ and $\sigma^*$, $x^*$ can be straightforwardly deduced for the original problem (\textbf{P1}) as shown in the proof of Theorem 1. 
\begin{algorithm}[!]
	\caption{Proposed hybrid algorithm for (\textbf{P1})}\label{euclid}
	\hspace*{\algorithmicindent} \textbf{Input:} Optimization problem (\textbf{P1}) \\
	\hspace*{\algorithmicindent} \textbf{Output:} $\boldsymbol{\eta}^*$ 
	\begin{algorithmic}[1]
		\State $\sigma \gets S$.
		\State Solve~\eqref{systemeq} to obtain $\boldsymbol{\eta}'$
		\If{$\boldsymbol{\eta}'$ satisfies~\eqref{p2c1}}    
		\State $\boldsymbol{\eta}^* \gets \boldsymbol{\eta}', \varsigma^* \gets S$, stop algorithm.
		\Else
		\State $high \gets S-1, low\gets 2 \sigma' \gets (high+low)/2$
		\Repeat
		\State Solve~\eqref{systemeq} to obtain $\boldsymbol{\eta}'$
		\If{$\boldsymbol{\eta}'$ satisfies~\eqref{p2c1}}         
		\State $low\gets \sigma'$, $\sigma' \gets (high+low)/2$
		\State $\varsigma^*=\sigma',\boldsymbol{\eta}^* \gets \boldsymbol{\eta}'$
		\Else
		\State $high\gets \sigma'$, $\sigma' \gets (high+low)/2$
		\EndIf
		\Until{$\sigma'=high$}
		\EndIf
		
	\end{algorithmic}
\end{algorithm}

\subsubsection{Optimality analysis}
In Lemma 1, we first prove that the solution obtained from solving~\eqref{systemeq} is the global optimal solution of (\textbf{P3}).
\begin{lemma}
Let $\boldsymbol{\eta}'$ denote a solution of~\eqref{systemeq}. $\boldsymbol{\eta}'$ is also the global optimal solution of (\textbf{P3}).
\end{lemma}
\begin{IEEEproof}
We will prove that $\boldsymbol{\eta}'$ satisfies the Karush–Kuhn–Tucker (KKT) conditions for non-convex optimization problems~\cite{complex1}. Moreover, since~\eqref{p3obj} and~\eqref{p3c1} are differentiable and satisfy linearity constraint qualification, strong duality holds, and thus $\boldsymbol{\eta}'$ is the global optimal solution of (\textbf{P3}). Next, we prove that $\boldsymbol{\eta}'$ satisfies the KKT conditions as follows. The first condition is:
\begin{equation}
f_i(\boldsymbol{\eta}')\leq 0.
\end{equation}
This condition is always satisfied since there is no inequality constraint ($f_i(\cdot)$) in (\textbf{P3}). The second condition is:
\begin{equation}
	\label{eq_second}
	h_k(\boldsymbol{\eta}')= 0=\sum_{s=1}^{\sigma} (\eta_k^s) -\eta_k, \forall k \in \mathcal{N}.
\end{equation}
The second condition is always satisfied since it is included in~\eqref{systemeq}. The third condition is
\begin{equation}
	\lambda_k\geq 0, \forall k \in \mathcal{N}.
\end{equation}
From~\eqref{systemeq}, we have
\begin{equation}
	\begin{split}
	\lambda_k&=-\ln({\tau})\sum_{n=1}^{N} (\eta_n^s)-2p^A_k\sum_{n=1}^N (0.5-p^A_n) \eta^s_n,\\
	&=\big(-\ln({\tau})-p^A_k)\big)\sum_{n=1}^{N} (\eta_n^s)+2p^A_k\sum_{n=1}^{N} (\eta_n^s).
	\end{split}
\end{equation}
Since $\tau \leq 0.001$ and $p_k^A<1$, we have $\lambda_k >0$, and thus the third condition is satisfied. The fourth condition is:
\begin{equation}
	\lambda_kh_k(\boldsymbol{\eta}')= 0.
\end{equation}
Similar to~\eqref{eq_second}, this condition is always satisfied since there is no inequality constraint in (\textbf{P3}). The fifth condition is 
\begin{equation}
	\begin{split}
		\nabla f_o(\boldsymbol{\eta}') + \sum_{k=1}^N\lambda_k \nabla h_k(\boldsymbol{\eta}')=0, \\
		=\lambda_k+\ln({\tau})\sum_{n=1}^{N} (\eta_n^s)+2p^A_k\sum_{n=1}^N (0.5-p^A_n) \eta^s_n=0,\\
		\forall k \in \mathcal{N}, \forall s= 1,\ldots \sigma,
	\end{split}
\end{equation}
which is included in~\eqref{systemeq}. As a result, $\boldsymbol{\eta}'$ satisfies all KKT conditions, and thus the proof is completed.
\end{IEEEproof}
Then, in Lemma 2, we prove that for any given $\sigma$, if the solution obtained from~\eqref{systemeq} satisfies~\eqref{p2c1}, it is the global optimal solution of (\textbf{P2}).

\begin{lemma}
		If the solution $\boldsymbol{\eta}'$ obtained from~\eqref{systemeq} satisfies~\eqref{p2c1}, $\boldsymbol{\eta}'$ is the global optimal solution of (\textbf{P2}).
\end{lemma}
\begin{IEEEproof}
	It follows from Lemma 1 that $\boldsymbol{\eta}'$ is the global optimal solution of (\textbf{P3}), and thus it satisfies~\eqref{p3c1}. Moreover, constraints~\eqref{p3c1} and~\eqref{p2c2} are identical. Therefore, $\boldsymbol{\eta}'$ satisfies~\eqref{p2c2}. Furthermore, the objective function~\eqref{p2obj} of (\textbf{P2}) is constant. As a result, if $\boldsymbol{\eta}'$ satisfies~\eqref{p2c1}, $\boldsymbol{\eta}'$ is the global optimal solution of (\textbf{P2}). The proof is completed.
\end{IEEEproof}
Next, in Theorem~\ref{theorem_equi}, we prove that for any given $\sigma$, if the solution $\boldsymbol{\eta}'$ obtained from solving~\eqref{systemeq} satisfies~\eqref{p2c1}, then we can straightforwardly derive an equivalent feasible solution of (\textbf{P1}). Moreover, if the optimal solution of (\textbf{P3}) satisfies~\eqref{p2c1} in the case where $\sigma= S$, we can derive the global optimal solution of (\textbf{P1}). Note that when the optimal solution of (\textbf{P3}) cannot satisfy~\eqref{p3obj}, it does not imply the absence of a feasible solution of (\textbf{P1}) for a given $\sigma$.  Despite this limitation, the proposed Lagrangian method can still find solutions that are better than those from commercial solvers in a significantly shorter amount of time. Moreover, the proposed method can find the global optimal solution in most experiments as later shown in Section~\ref{simu}. 
\begin{theorem}
	\label{theorem_equi}
	For any given $\sigma$, if the solution $\boldsymbol{\eta}'$ obtained from~\eqref{systemeq} satisfies~\eqref{p2c1}, then $\{\boldsymbol{\eta}',\mathbf{x},\sigma\}$ is a feasible solution to (\textbf{P1}), where $\mathbf{x}$ can be straightforwardly derived from $\sigma$.
\end{theorem}
\begin{IEEEproof}
First, we prove that for any specific $\sigma$, we can straightforwardly derive $\mathbf{x}$. Substituting $\varsigma=\sigma$ into~\eqref{c2}, we have
\begin{equation}
	x_s \geq \dfrac{\sigma-s+1}{S}, \forall s=1,\ldots, S.
\end{equation}
This means that $x_s>0$, $\forall s\leq \sigma$. Then, substituting $\varsigma=\sigma$ into~\eqref{c3}, we have
\begin{equation}
	x_s \leq \sigma-s+1, \forall s=1,\ldots, S.
\end{equation}
This means that $x_s \leq 0$, $\forall s> \sigma$. Moreover, since $\mathbf{x}$ are binary, we have $x_s=1$ , $\forall s=1, \ldots, \sigma$ and $x_s=0$, $\forall s> \sigma$.

As a result,~\eqref{c1} becomes
\begin{equation}
	\label{eq_proof1}
\bigg(\sum_{n=1}^N (0.5-p^A_n) \eta^s_n\bigg)^2\geq -0.5\ln(\tau)\sum_{n=1}^{N} (\eta_n^s)^2, \forall s \leq \sigma,
\end{equation}
and
\begin{equation}
		\label{eq_proof2}
	\bigg(\sum_{n=1}^N (0.5-p^A_n) \eta^s_n\bigg)^2\geq 0, \forall s > \sigma.
\end{equation}
Since $\boldsymbol{\eta}'$ satisfies~\eqref{p2c1}, it also satisfies~\eqref{eq_proof1}. Moreover, since $p^A_n<0.5$ and $\eta_n^s >0$,~\eqref{eq_proof2} is always satisfied. As a result, $\{\boldsymbol{\eta}',\mathbf{x},\sigma\}$ satisfy all constraints of (\textbf{P1}), and thus it is a feasible solution to (\textbf{P1}). The proof is now completed.
\end{IEEEproof}
\subsubsection{Complexity analysis}
The main component of Algorithm 1 is solving~\eqref{systemeq} to obtain $\boldsymbol{\eta}'$ in Steps 2 and 8. Using methods such as Gaussian elimination~\cite{complex1}, each instance of~\eqref{systemeq} can be solved with time complexity $O\big((\sigma N+N)^3\big)$. Additionally, because we utilize binary search,~\eqref{systemeq} needs to be solved at most $\textrm{log}(S)$ times, and thus the total time complexity of Algorithm 1 is $O\big(\textrm{log}(S) (\sigma N+ N)^3\big)$. In contrast, the time complexity of solving (\textbf{P1}) is exponential~\cite{complex1}, and (\textbf{P1}) involves more variables. As a result, Algorithm 1 can be frequently deployed to reconfigure the shards, thereby reducing the risk of corruption from the adversary.


			\begin{figure*}[!t]
	\includegraphics[width=.7\textwidth]{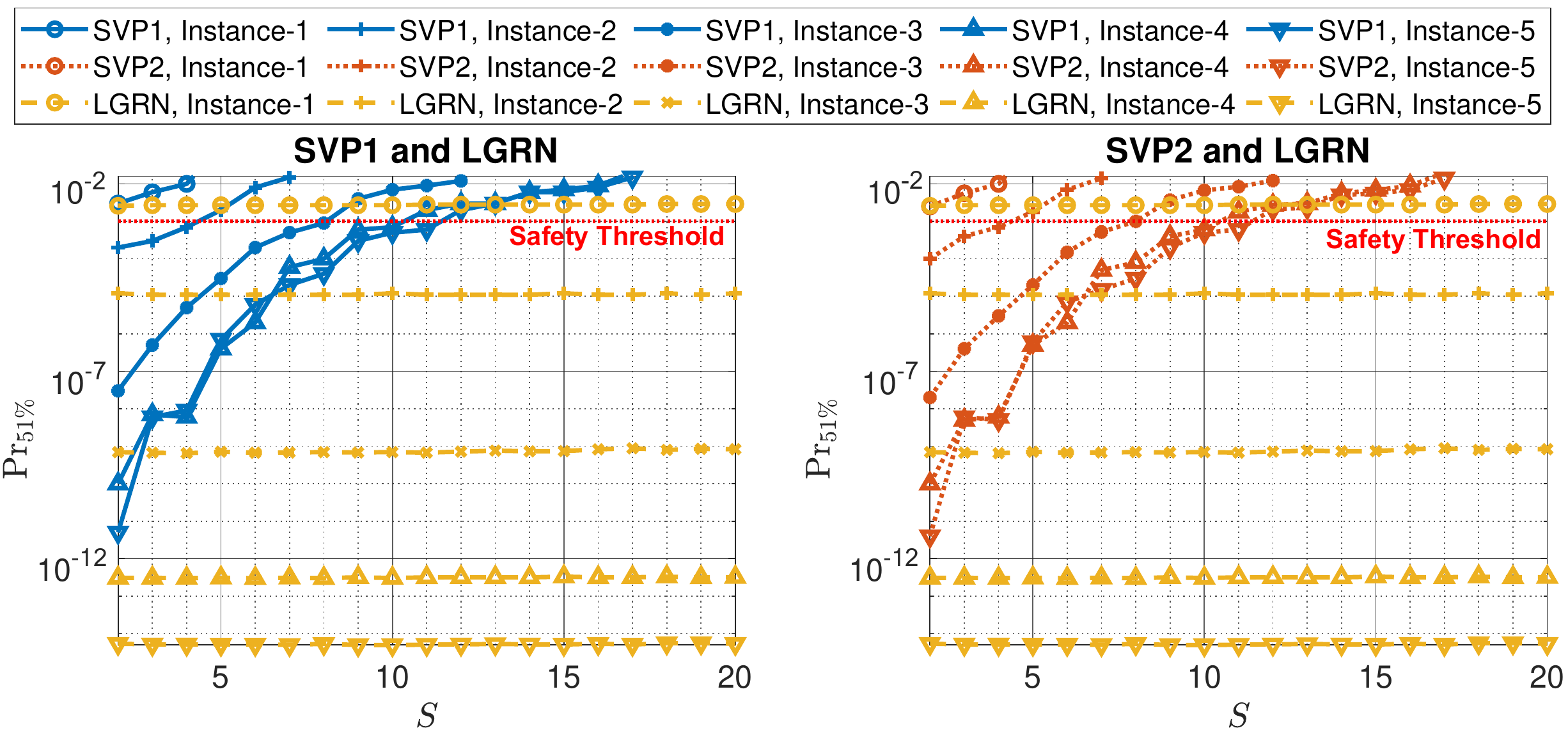}
	\centering
	\caption{$\rm{Pr}_{51\%}$ achieved by the three methods.}
	\label{Fig:exp1}
\end{figure*}

\section{Performance Evaluation}
\label{simu} 
\subsection{Simulation Settings}
To evaluate the performance of our proposed approach, we conduct various numerical experiments in five problem instances with different parameters (number of nodes, maximum difference, mean, and standard deviation (STD) of MUs score distribution) as shown in Table~\ref{tab:tab1}. Moreover, in all experiments, we set $T=2000$ Tx/s and $\alpha_C=\alpha_D=\alpha_T=1$. In these experiments, we compare the performance of three methods as follows:
\begin{itemize}
	\item $SVP1$: We solve (\textbf{P1}) directly using the commercial solver CPLEX~\cite{solver}.
	\item $SVP2$: We apply an iterative algorithm similar to Algorithm 1. However, instead of solving~\eqref{systemeq} in Steps 2 and 8 as done in Algorithm 1, we solve (\textbf{P2}) using the commercial solver CPLEX~\cite{solver}.
	\item $LGRN$: We solve (\textbf{P3}) using the proposed Lagrangian approach as described in Algorithm 1.
\end{itemize}

\begin{table}[!]
	\footnotesize
	\centering
	\caption{Problem instance parameters.} 
	\begin{tabular}{|>{\raggedright\arraybackslash}m{1cm}|>{\raggedright\arraybackslash}m{1cm}|>{\raggedright\arraybackslash}m{1.5cm}|>{\raggedright\arraybackslash}m{1cm}|>{\raggedright\arraybackslash}m{1cm}|}
		\hline 
		{\centering\arraybackslash}{\textbf{Instance}} &
		{\centering\arraybackslash}{\textbf{No. of nodes}} &
		{\centering\arraybackslash}{\textbf{Maximum difference }}&
		{\centering\arraybackslash}{\textbf{Mean}}&
		{\centering\arraybackslash}{\textbf{STD}}\\
		\hline 
		\hline
		1          &25   & 29   & 39.0  &       7.9          \\ 
		\hline
		2     &50        & 31             & 36.8 & 6.7\\ 
		\hline
		3     &100     &     38           & 38.4         &  4.8                \\ 
		\hline
		4   &150     & 109 &  89.9      & 19.9  \\ 
		\hline
		5&200 &   170                    &  123.8      & 32.9  \\
		\hline
	\end{tabular}
	\label{tab:tab1}
\end{table}

In the first set of experiments, we examine the best solution found by the three methods under a limited running time (1 minute). Particularly, for each instance, we vary $\varsigma$ and $\tau$ to examine the best possible solution found by each method. The results show the lowest probability that the adversary can control more than 51\% of a shard's score, denoted by $\rm{Pr}_{51\%}$ ($\rm{Pr}_{51\%}$ can be calculated using~\eqref{prob_1}). For each method, we record the lowest $\rm{Pr}_{51\%}$ given a specific number of shards.

In the second set of experiments, we let all three methods run up to 10 minutes and then compare their running time and achievable throughput. For $SVP2$, we set the time limit of each iteration (Step 2 and Step 8 in Algorithm 1) to 1 minute. Moreover, we conduct experiments with different values of $S$ to show the impact of $S$ on the performance of the considered methods.

In the third set of experiments, we vary the values of $p_n^A$ to study the impacts of the adversarial probability on the performance and security of the network. Particularly, we gradually increase $p_n^A$ and examine the best achieved $\rm{Pr}_{51\%}$ of the three methods for various numbers of shards. Moreover, we also measure the highest throughput achieved by the considered methods.

Finally, we study the impact of the MUs' scores on the security of the system. In particular, for a network of 50 nodes, we randomly generate instances with different user engagement scores, as reflected by the different standard deviations and average values of engagement scores. Then, for each instance, we examine the best $\rm{Pr}_{51\%}$ achieved by the proposed $LGRN$ method to study how different distributions of scores can affect network security. 
\subsection{Simulation Results}

Fig.~\ref{Fig:exp1} illustrates the best $\rm{Pr}_{51\%}$ obtained by the three methods for different numbers of shards in all problem instances. For example, in Instance-1 with 25 nodes, when we want to optimize the score allocation for 2 shards, the three methods achieve similar results, e.g., around 0.003 possibility to be attacked. However, if we want to have more shards in the system, $\rm{Pr}_{51\%}$ increases drastically if we use the $SVP1$ and $SVP2$ methods, e.g., 0.006 for 3 shards, 0.01 for 4 shards, and more than 0.01 for higher numbers of shards. In contrast, even for 20 shards, the value of $\rm{Pr}_{51\%}$ achieved by the $LGRN$ method is only around 0.003. Moreover, for all other instances, the $LGRN$ method can achieve $\rm{Pr}_{51\%}$ lower than the safety threshold (0.001) for up to 20 shards in the system. In contrast, $SVP1$ and $SVP2$ can only ensure security, i.e., $\rm{Pr}_{51\%} <0.001$, for up to 4, 8, 10, and 11 shards in instances 2, 3, 4, and 5, respectively. Furthermore, compared to $SVP1$ and $SVP2$, $LGRN$ can achieve smaller $\rm{Pr}_{51\%}$ in all cases. Note that since the values of $\rm{Pr}_{51\%}$ achieved by $LGRN$ does not vary much compared to the other methods, it is not shown clearly in the figure. For example, in Instance-3, the values of $\rm{Pr}_{51\%}$ achieved by $LGRN$ ranges from 6x$10^{-10}$ to 9x$10^{-10}$, whereas those achieved by $SVP1$ ranges from 3x$10^{-8}$ to 0.012.

			\begin{figure}[!]
	\includegraphics[width=.35\textwidth]{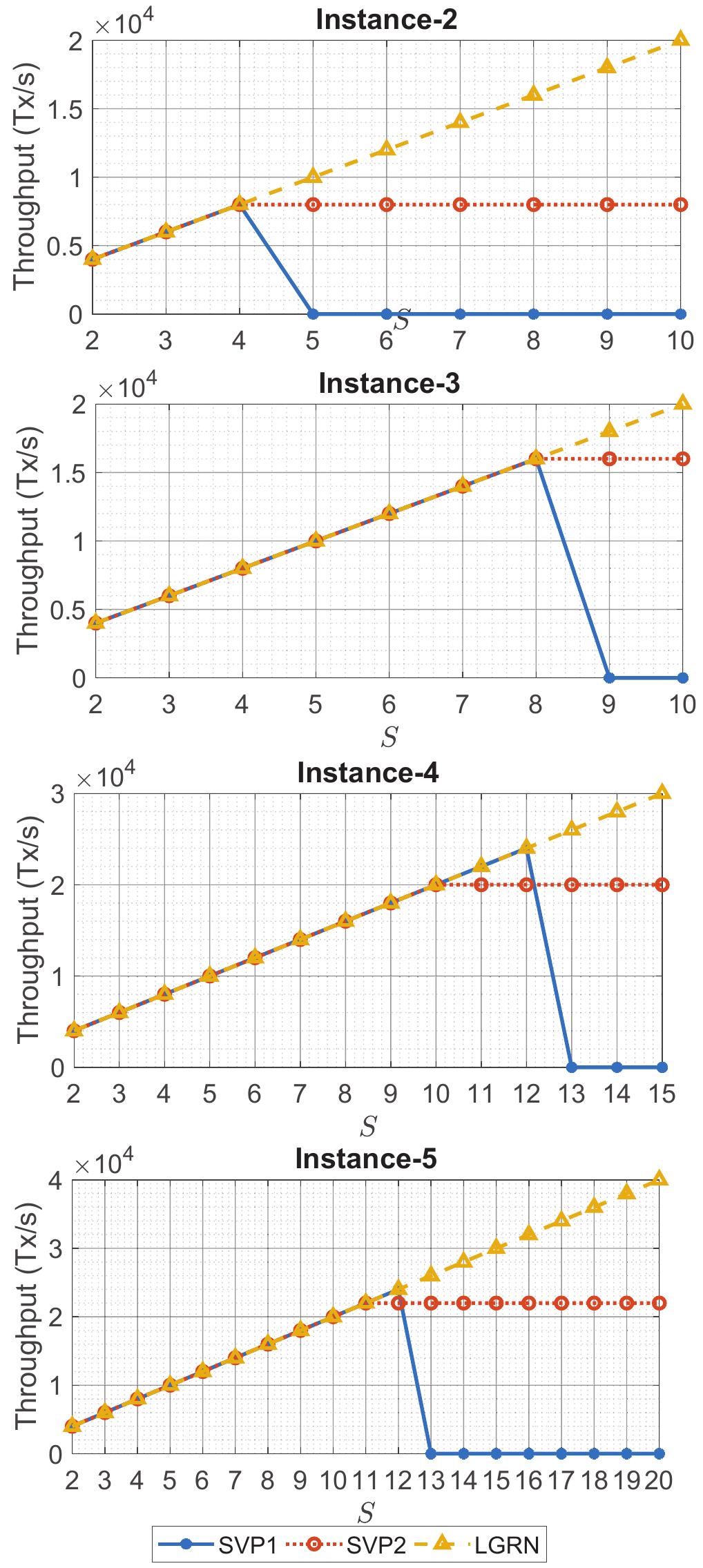}
	\centering
	\caption{Throughput achieved by the three methods.}
	\label{Fig:exp2_sol}
\end{figure}
Fig.~\ref{Fig:exp2_sol} shows the throughput achieved by the three methods for Instance-2 to Instance-5. We do not show the achieved throughput for Instance-1 because, in this instance, all three methods cannot ensure that $\rm{Pr}_{51\%}$ is lower than the safety threshold even for 2 shards, and thus the network cannot be divided into shards. For all the remaining instances, we can observe that the proposed $LGRN$ method performs better than the other methods, especially for high numbers of shards. For example, in Instance 2, $LGRN$ can achieve a throughput up to $20,000$ Tx/s, while the other methods can achieve at most $8,000$ Tx/s. Moreover, the $SVP1$ fails to find a feasible solution for $S>5$, and thus the network cannot be divided into shards in these cases. Similarly, $LGRN$ performs better than the other methods by up to $25\%$, $50\%$, and $66.6\%$, in Instances 3, 4, and 5, respectively. Moreover, in Instance-2 to Instance-5, $LGRN$ can achieve global optimal solutions for all values of $S$, whereas $SVP1$ and $SVP2$ can not for higher values of $S$. 
			\begin{figure}[!]
	\includegraphics[width=.35\textwidth]{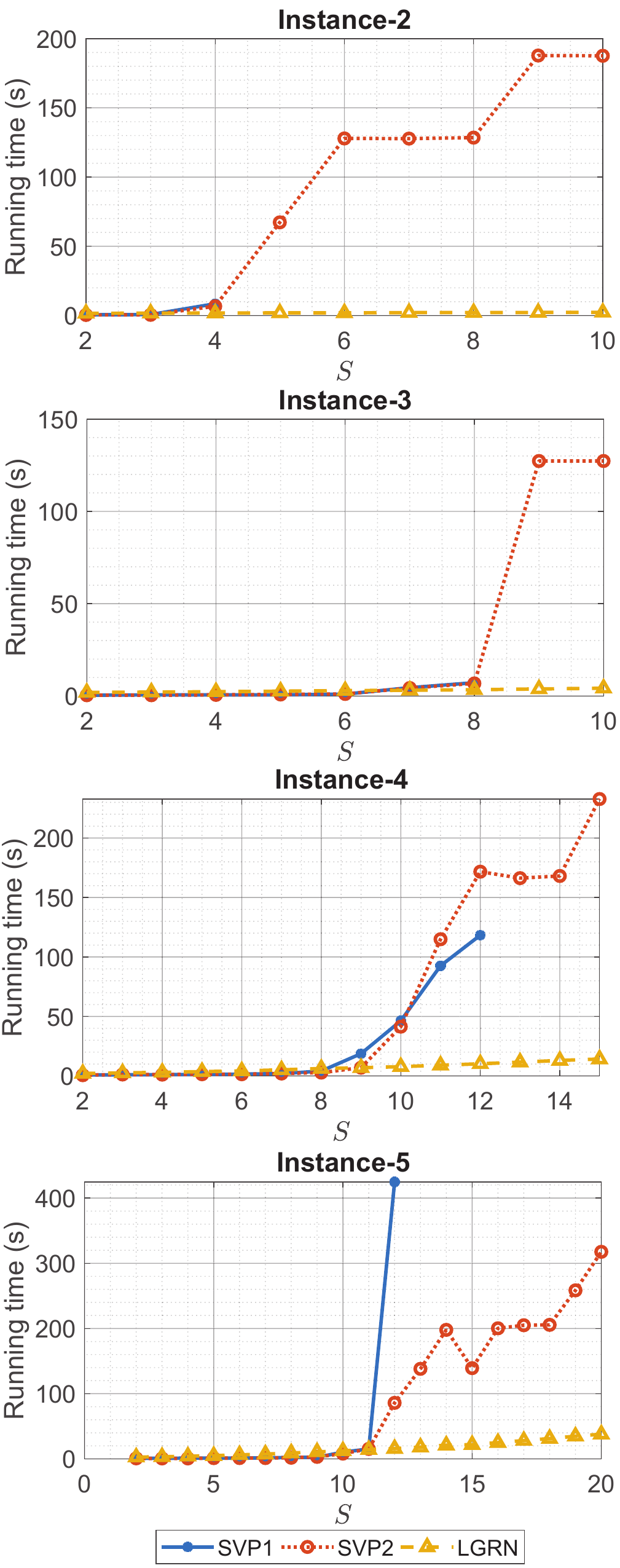}
	\centering
	\caption{Running time of the three methods.}
	\label{Fig:exp2_time}
\end{figure}

Fig.~\ref{Fig:exp2_time} shows the running time of the three methods in Instance-2 to Instance-5. As observed, the computational time of the proposed $LGRN$ is trivial compared to the other methods, particularly for high numbers of shards. For example, in Instance 2, $LGRN$ needs only 2.3 seconds to find the solution to divide the network into 10 shards. In contrast, $SVP2$ needs more than 187 seconds, whereas $SVP1$ can only find solutions for up to 4 shards. For more than 5 shards, $SVP1$ exceeds the running time limit without being able to find any feasible solution. Similarly, for the remaining instances, $LGRN$ can find better solutions in a much shorter time, i.e., more than 30, 16, and 8 times faster than $SVP2$. Meanwhile, $SVP1$ fails to find any feasible solution for 9, 13, and 12 shards in Instance-3, Instance-4, and Instance-5, respectively. Because of that, the graphs in Fig.~\ref{Fig:exp2_time} do not show the running time of $SVP1$ in the cases where $SVP1$ cannot find any feasible solution. Additionally, we can observe that the running time of $LGRN$ scales almost linearly with $S$, while the running time of the other methods increases exponentially as $S$ increases.
			\begin{figure}[!]
	\includegraphics[width=.4\textwidth]{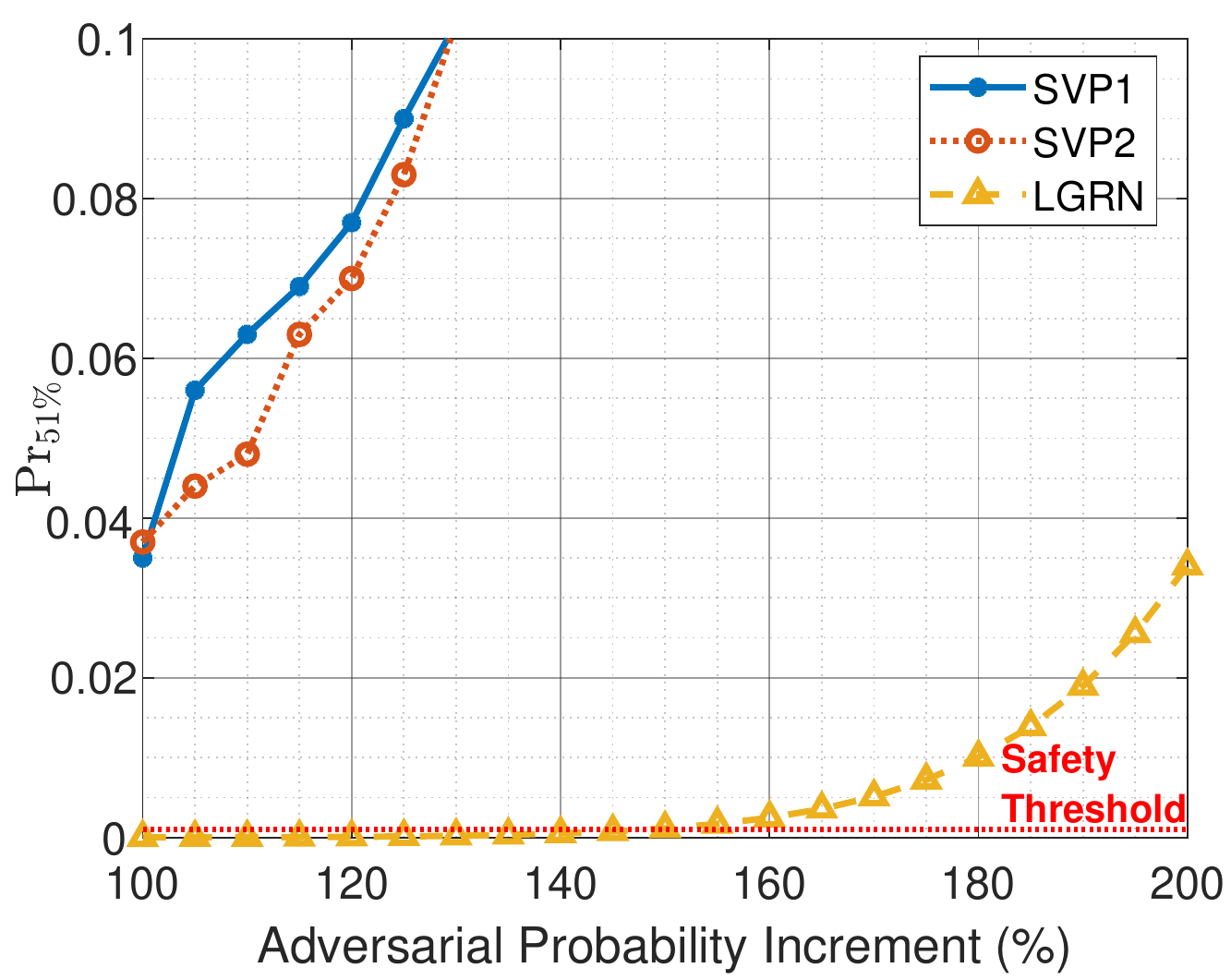}
	\centering
	\caption{$\rm{Pr}_{51\%}$ under increasing adversarial probability.}
	\label{Fig:exp3_prob}
\end{figure}

Fig.~\ref{Fig:exp3_prob} illustrates the results of the third set of experiments in terms of security, i.e., the change in $\rm{Pr}_{51\%}$ as the adversarial probability ($p_n^A$) increases. As observed from the figure, $LGRN$ can ensure the safety ($\rm{Pr}_{51\%}<0.1\%$) of a network with 50 nodes and 10 shards even when $p_n^A$ increases by nearly 150\%. In contrast, if we use $SVP1$ and $SVP2$, $\rm{Pr}_{51\%}$ is nearly 0.04. Moreover, as $p_n^A$ increases, $\rm{Pr}_{51\%}$ obtained from $SVP1$ and $SVP2$ increases drastically to over 0.1. This means that these methods cannot be employed for sharding when the adversary controls a high portion of MUs. Furthermore, we can also observe that $SVP2$ performs slightly better than $SVP1$ as the adversarial probability increases.
			\begin{figure}[!]
	\includegraphics[width=.4\textwidth]{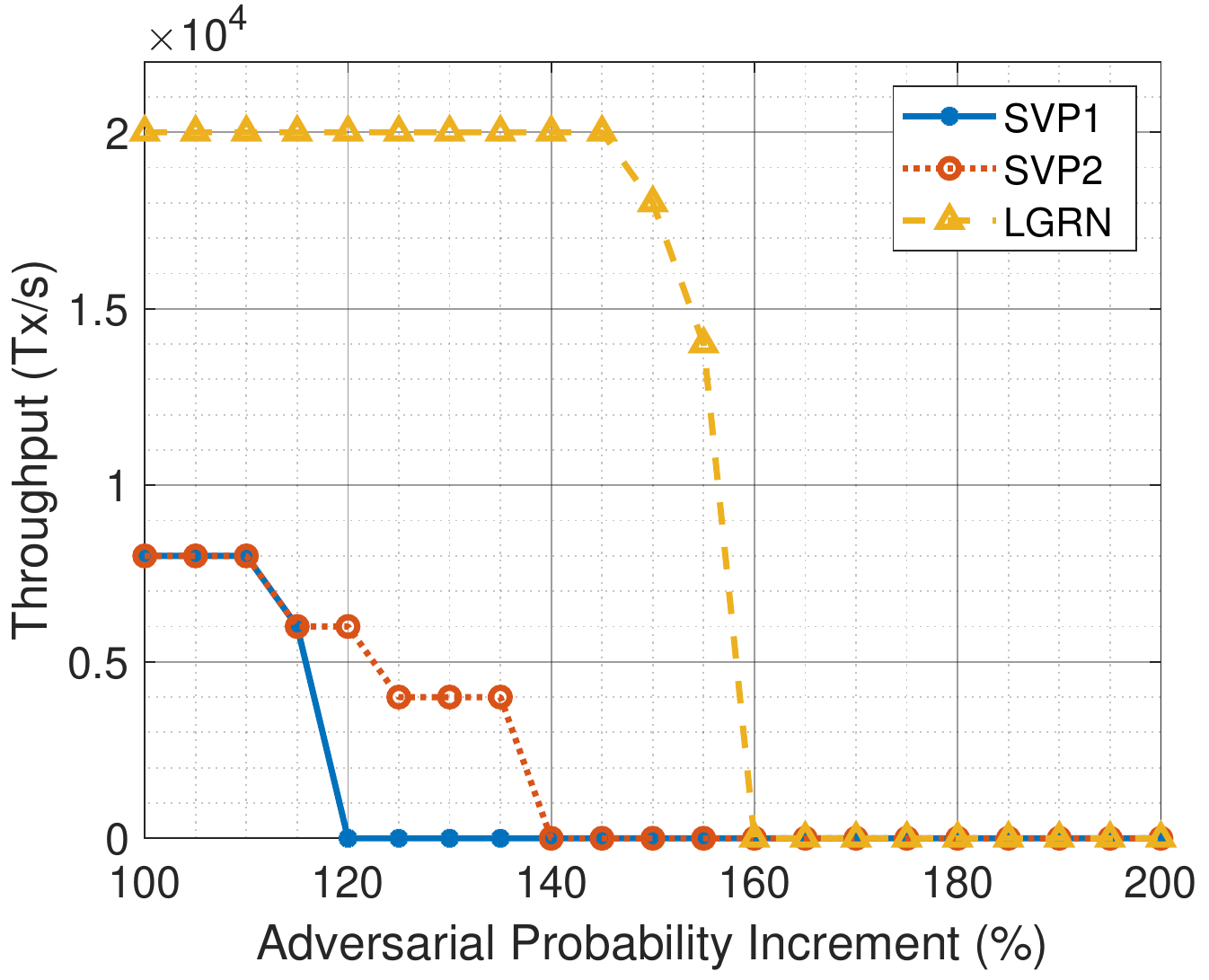}
	\centering
	\caption{Throughput under increasing adversarial probability.}
	\label{Fig:exp3_thru}
\end{figure}

Fig.~\ref{Fig:exp3_thru} shows the highest throughput achieved by the three methods as $p_n^A$ increases. It can be observed that the highest throughput $LGRN$ can achieve is 20,000 Tx/s (global optimal) with up to 145\% increase in adversarial probability. In contrast, $SVP1$ and $SVP2$ only attain a maximum of 8,000 Tx/s and their throughput decreases when $p_n^A$ exceeds 115\%. When $p_n^A$ continues to rise, $SVP1$ and $SVP2$ fail to divide the network into shards at 120\% and 140\% respectively, while $LGRN$ can still sustain a throughput of 14,000 Tx/s at 155\%. $LGRN$ only fails to find a feasible solution after $p_n^A$ increases by more than 160\%.   
			\begin{figure}[!]
	\includegraphics[width=.45\textwidth]{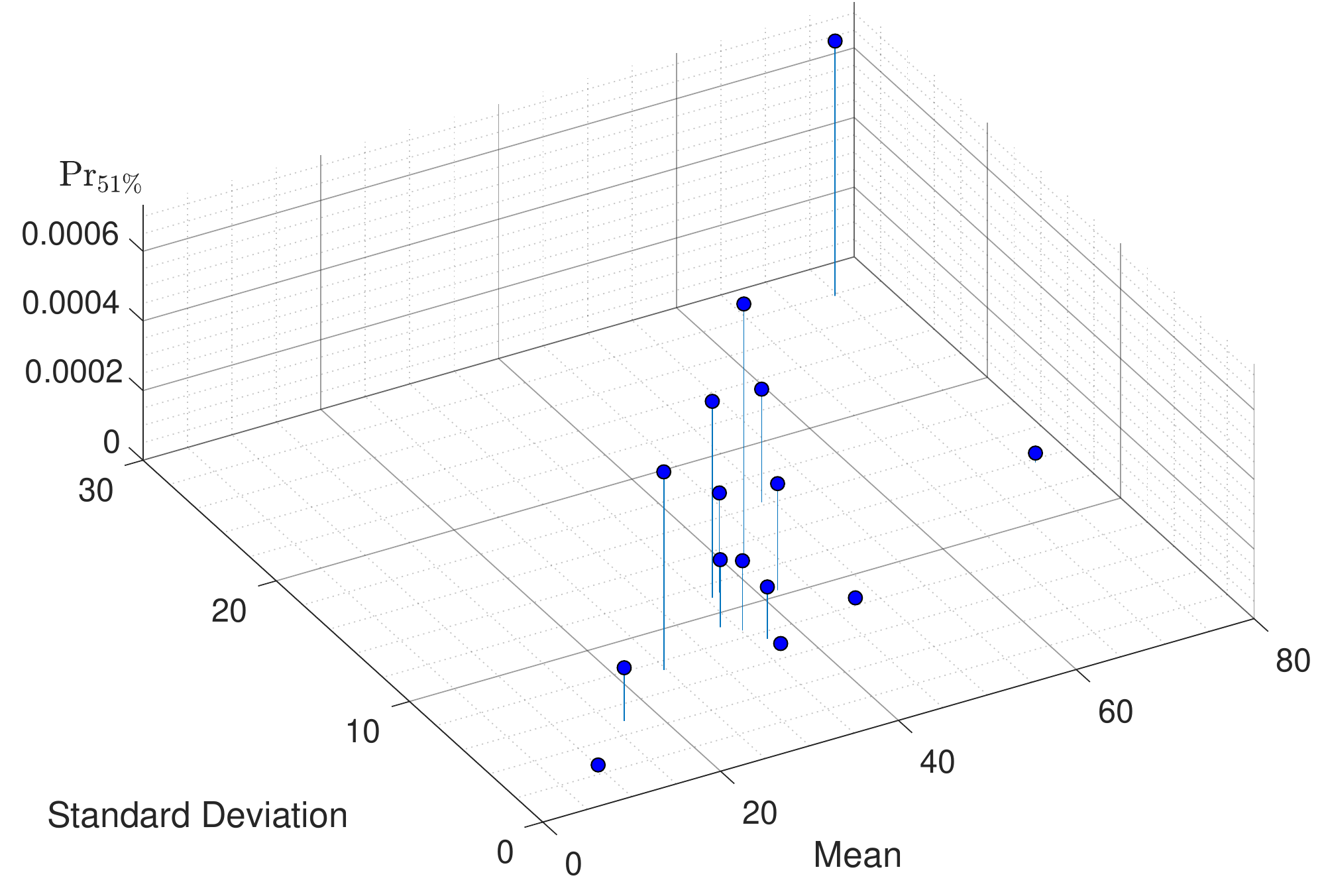}
	\centering
	\caption{Impacts of mean and standard deviation.}
	\label{Fig:exp4}
\end{figure}

Fig.~\ref{Fig:exp4} illustrates the $\rm{Pr}_{51\%}$ achieved by $LGRN$ for different distributions of engagement scores. As observed from the figure, the more spread out the engagement scores are (i.e., the standard deviation is high), the higher the possibility of the network being attacked by the adversary. Moreover, the higher score the network has (i.e., higher mean), the more secure it becomes. For example, the instance with the largest standard deviation has the greatest likelihood of being attacked. Moreover, among instances with similar standard deviations, the ones with a higher mean (which corresponds to a higher total score) have a lower probability of being attacked. Therefore, while the network operators cannot influence the score distribution, they can try to attract more MUs to the network (thereby increasing the total score) to improve network security and performance.
\section{Conclusion}
\label{conclu}
In this paper, we have developed a novel sharding blockchain framework for Metaverse applications. Particularly, we have developed a PoE consensus mechanism that can encourage and reward MUs' resources contribution, thereby alleviating the huge resource demands for MSP and creating a more engaged MU community. Moreover, we have proposed a sharding management scheme and formulated an optimization problem to find the optimal number of shards and MUs allocation. Since the optimization problem is NP-complete, we have developed a hybrid approach that decomposes the problem (using the binary search method) into sub-problems that can be solved effectively by the Lagrangian method. As a result, the proposed approach can obtain solutions in polynomial time, thereby enabling flexible shard reconfiguration and reducing the risk of corruption from the adversary. Extensive numerical experiments have been conducted, and their results have shown that, compared to the state-of-the-art commercial solvers, our proposed approach can achieve up to 66.6\% higher throughput in less than 1/30 running time. Moreover, the proposed approach can achieve global optimal solutions in most experiments. Furthermore, we have studied the impacts of key parameters on the performance of the system and shown that the proposed approach can further improve the robustness of the system. 
	\bibliographystyle{IEEE}

			\begin{IEEEbiography}[{\includegraphics[width=1in,height=1.25in,clip,keepaspectratio]{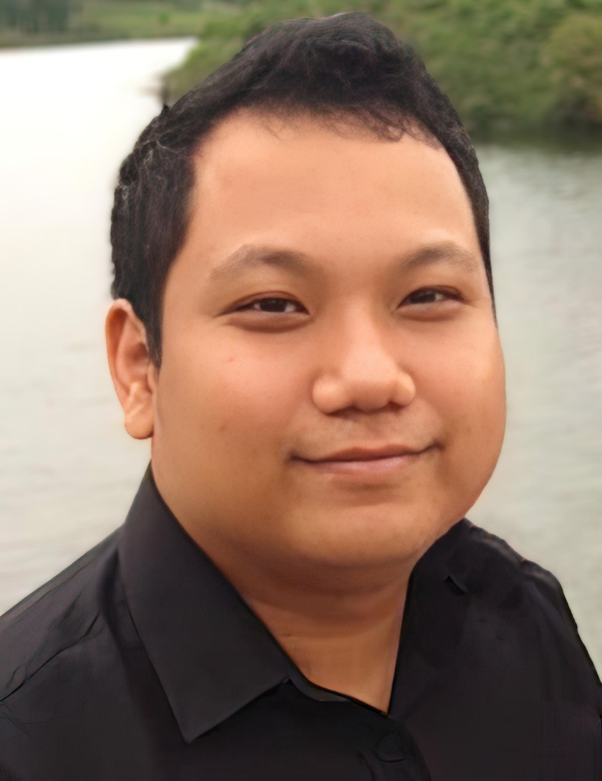}}]{Cong T. Nguyen} received his B.E. degree in Electrical Engineering and Information from the Frankfurt University of Applied Sciences in 2014, his M.Sc. degree in Global Production Engineering and Management from the Technical University of Berlin in 2016. Since 2019, he has been a Ph.D. student at the  UTS-HCMUT Joint Technology and Innovation Research Centre  between Ho Chi Minh University of Technology and the University of Technology Sydney (UTS). His research areas include operations research, blockchain technology, game theory and optimizations. 
\end{IEEEbiography}
\begin{IEEEbiography}[{\includegraphics[width=1in,height=1.25in,clip,keepaspectratio]{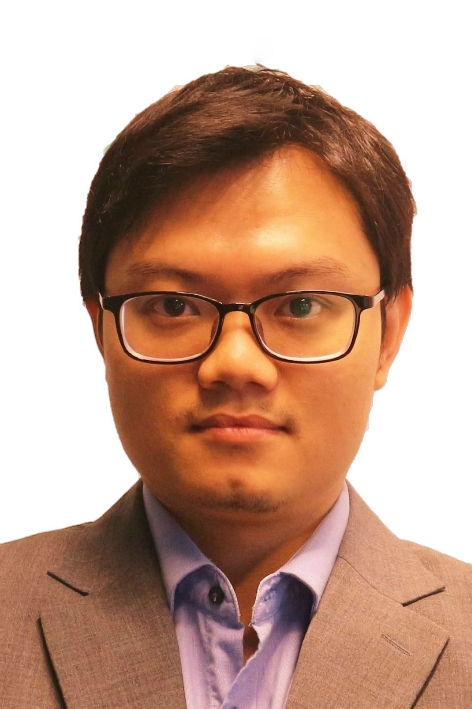}}]{Dinh Thai Hoang} (M'16-SM'22) is currently a faculty member at the School of Electrical and Data Engineering, University of Technology Sydney, Australia. He received his Ph.D. in Computer Science and Engineering from the Nanyang Technological University, Singapore, in 2016. His research interests include emerging topics in wireless communications and networking such as machine learning, edge intelligence, cybersecurity, IoT, and Metaverse. He has received several awards including Australian Research Council and IEEE TCSC Award for Excellence in Scalable Computing (Early Career Researcher). Currently, he is an Editor of IEEE Transactions on Wireless Communications, IEEE Transactions on Cognitive Communications and Networking, IEEE Transactions on Vehicular Technology, and Associate Editor of IEEE Communications Surveys \& Tutorials. 
\end{IEEEbiography}
\begin{IEEEbiography}[{\includegraphics[width=1in,height=1.25in,clip,keepaspectratio]{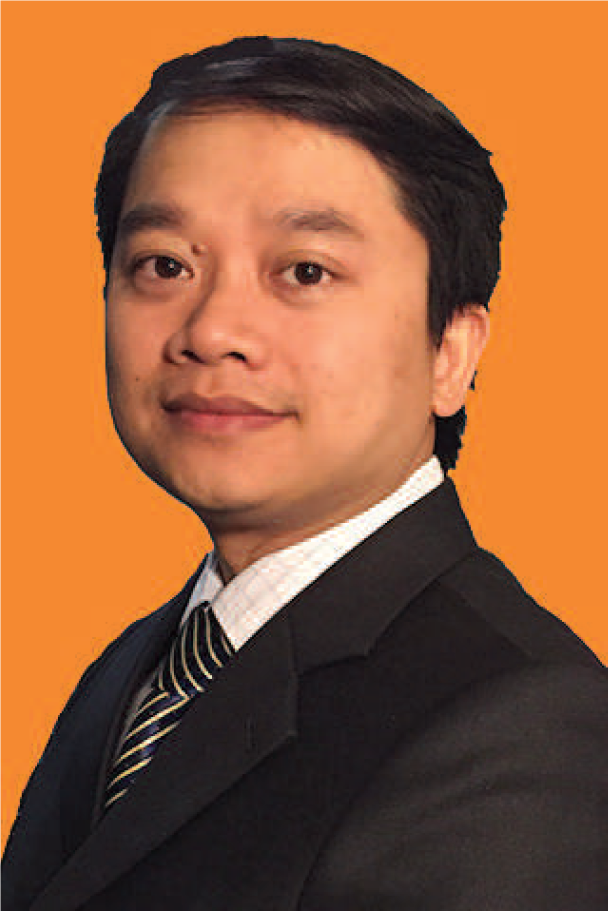}}]{Diep N. Nguyen} (M'13-SM'19) received the M.E. degree in electrical and computer engineering from the University of California at San Diego (UCSD) and the Ph.D. degree in electrical and computer engineering from The University of Arizona (UA). He is currently a Faculty Member with the Faculty of Engineering and Information Technology, University of Technology Sydney (UTS). Before joining UTS, he was a DECRA Research Fellow with Macquarie University and a Member of Technical Staff with Broadcom Corporation, Irvine, CA, USA, and ARCON Corporation, Boston, MA, USA, and consulting the Federal Administration of Aviation on turning detection of UAVs and aircraft, and the U.S. Air Force Research Laboratory on anti-jamming. His research interests include computer networking, wireless communications, and machine learning application, with emphasis on systems' performance and security/privacy. He received several awards from LG Electronics, UCSD, UA, the U.S. National Science Foundation, and the Australian Research Council. He is currently an Editor, Associate Editor, Guest Editor of the IEEE Transactions on Mobile Computing, IEEE Access, Sensors journal, IEEE Open Journal of the Communications Society (OJ-COMS), and Scientific Reports (Nature's).
\end{IEEEbiography}
\begin{IEEEbiography}[{\includegraphics[width=1in,height=1.25in,clip,keepaspectratio]{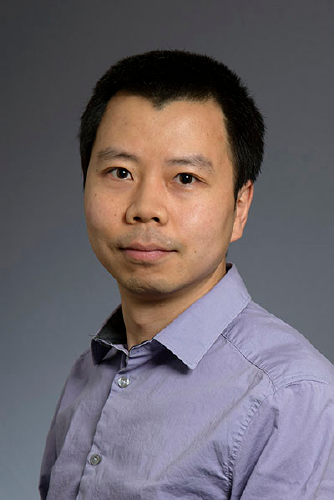}}]{Yong Xiao}(S'09-M'13-SM'15) received his B.S. degree in electrical engineering from China University of Geosciences, Wuhan, China in 2002, M.Sc. degree in telecommunication from Hong Kong University of Science and Technology in 2006, and his Ph. D degree in electrical and electronic engineering from Nanyang Technological University, Singapore in 2012. He is now a professor in the School of Electronic Information and Communications at the Huazhong University of Science and Technology (HUST), Wuhan, China. He is also with Peng Cheng Laboratory, Shenzhen, China and Pazhou Laboratory (Huangpu), Guangzhou, China. He is the associate group leader of the network intelligence group of IMT-2030 (6G promoting group) and the vice director of 5G Verticals Innovation Laboratory at HUST. Before he joins HUST, he was a research assistant professor in the Department of Electrical and Computer Engineering at the University of Arizona where he was also the center manager of the Broadband Wireless Access and Applications Center (BWAC), an NSF Industry/University Cooperative Research Center (I/UCRC) led by the University of Arizona. His research interests include machine learning, game theory, distributed optimization, and their applications in semantic communications, semantic-aware networks, cloud/fog/mobile edge computing, green communication systems, and Internet-of-Things (IoT).
\end{IEEEbiography}

\begin{IEEEbiography}[{\includegraphics[width=1in,height=1.25in,clip,keepaspectratio]{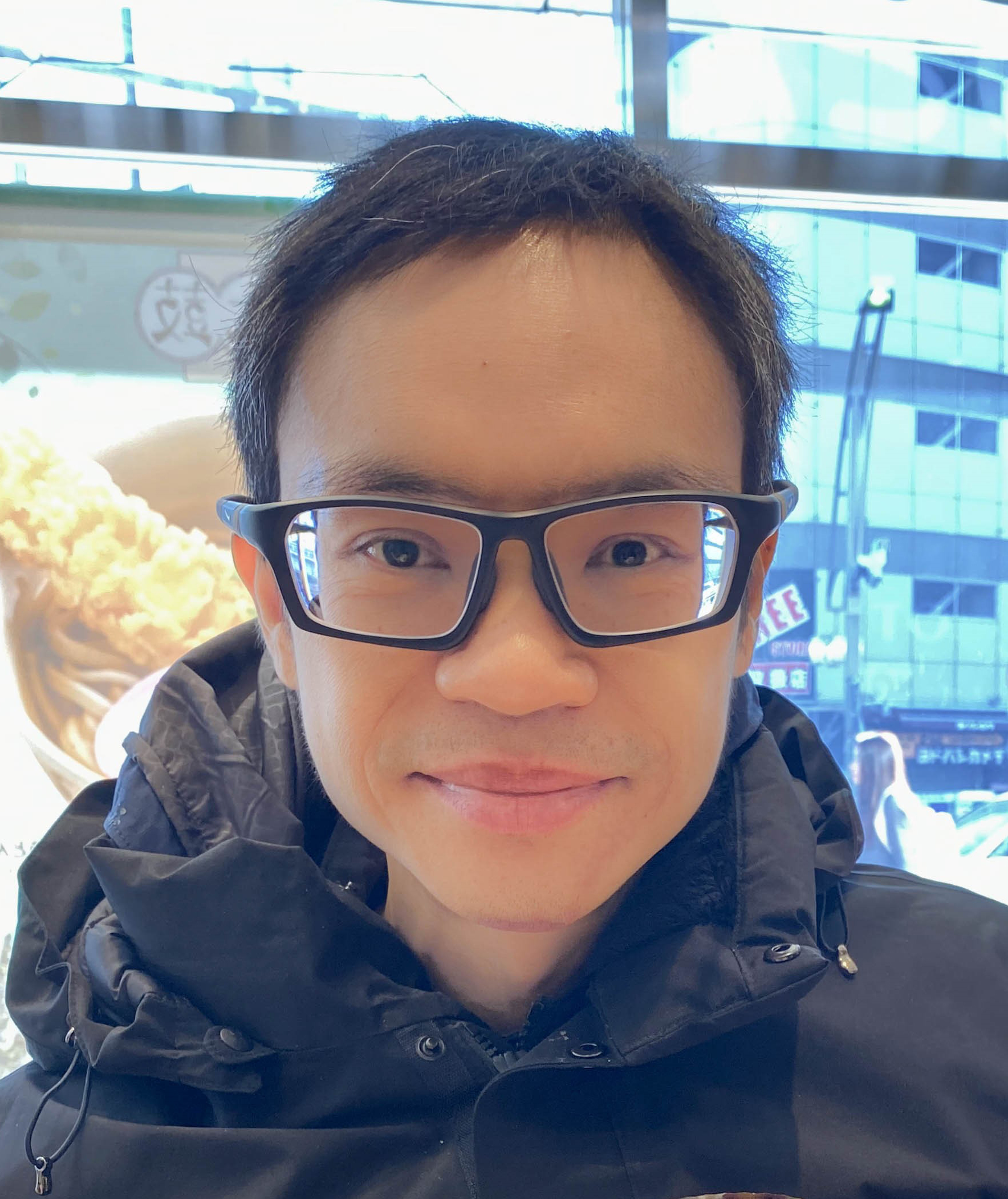}}]{Dusit Niyato} (M'09-SM'15-F'17) is a professor in the School of Computer Science and Engineering, at Nanyang Technological University, Singapore. He received B.Eng. from King Mongkuts Institute of Technology Ladkrabang (KMITL), Thailand in 1999 and Ph.D. in Electrical and Computer Engineering from the University of Manitoba, Canada in 2008. His research interests are in the areas of Internet of Things (IoT), machine learning, and incentive mechanism design.

\end{IEEEbiography}

\begin{IEEEbiography}[{\includegraphics[width=1in,height=1.25in,clip,keepaspectratio]{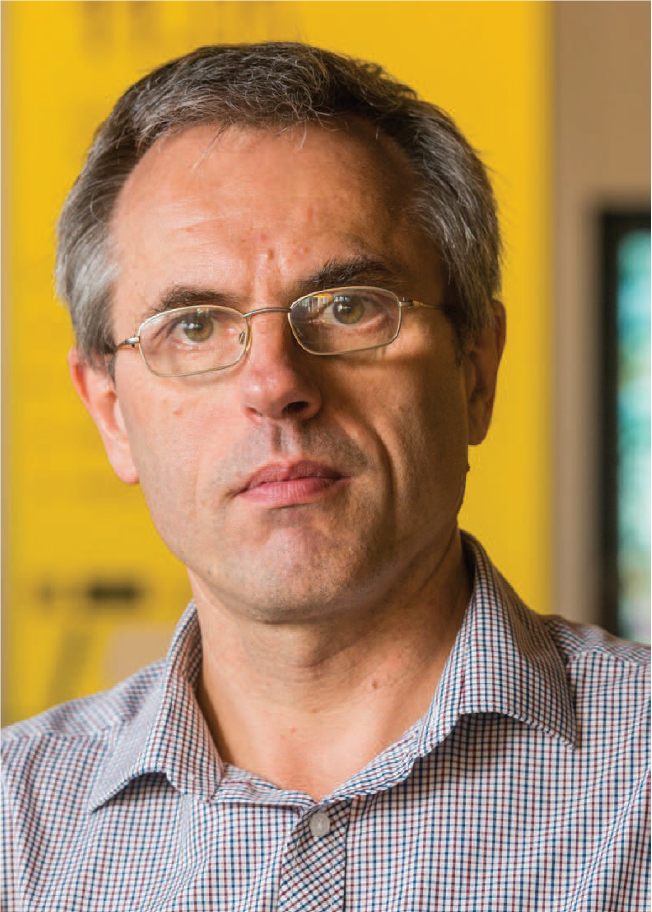}}]{Eryk Dutkiewicz} (M'05--SM'15) received his B.E. degree in Electrical and Electronic Engineering from the University of Adelaide in 1988, his M.Sc. degree in Applied Mathematics from the University of Adelaide in 1992 and his PhD in Telecommunications from the University of Wollongong in 1996. His industry experience includes management of the  Wireless Research Laboratory at Motorola in early 2000's. Prof. Dutkiewicz is currently the Head of School of Electrical and Data Engineering at the University of Technology Sydney, Australia. He is a Senior Member of IEEE. He also holds a professorial appointment at Hokkaido University in Japan. His current research interests cover 5G/6G and IoT networks. 
\end{IEEEbiography}

\end{document}